\documentclass[fleqn,usenatbib]{mnras}
\usepackage[T1]{fontenc}
\usepackage{graphicx}
\usepackage{amsmath}
\usepackage{amssymb}

\usepackage{epstopdf}

\usepackage{times}
\usepackage{amsfonts}
\usepackage{hyperref}

\title [Li dispersion in PMS stars]{The growth of a lithium abundance dispersion in pre main sequence stars}
\author[R. J. Jackson et al.]
  {R. J.~Jackson$^1$, R.~D. Jeffries$^1$ and E.~Tognelli$^2$\\
  $^1$Astrophysics Group, Keele University, Keele, Staffordshire ST5 5BG, United Kingdom\\
$^2$University of New Haven, West Haven, Connecticut 06516, USA}

\date{27th September 2024}

\pagerange{\pageref{firstpage}--\pageref{lastpage}} \pubyear{2024}

\def\LaTeX{L\kern-.36em\raise.3ex\hbox{a}\kern-.15em
    T\kern-.1667em\lower.7ex\hbox{E}\kern-.125emX}

\begin{document}
\label{firstpage}
\maketitle

\begin{abstract}
Lithium is predicted, and observed, to be depleted in contracting, low-mass pre main sequence (PMS) stars. Yet these stars reach the zero age main sequence (ZAMS) with a spread in lithium abundance at a given effective temperature that is not predicted by standard stellar evolutionary models and which appears to be correlated with rotation. Using a homogeneous dataset provided by the {\it Gaia}-ESO spectroscopic survey, we have followed the evolving photospheric lithium content of cohorts of stars destined to be ZAMS late G-, K- and M-dwarfs, in clusters at ages of 2--300\,Myr. We show that a  dispersion in the Li~{\sc i}~6708\AA\ line strength develops in the lower mass stars after 10--20\,Myr on the PMS, as soon as Li depletion begins, even in fully convective stars. A model based on a surface starspot coverage varying from star-to-star, leading to a differential Li-burning rate, can explain this temporal behaviour and its mass dependence. However, to fully explain the magnitude of the Li dispersion and its correlation with rotation, the spot coverage during Li-burning would need to be a factor of two larger on average than measured in ZAMS clusters like the Pleiades and continue increasing with rotation in PMS stars beyond the usual "saturation limit" observed for other magnetic activity indicators.

\end{abstract}


\section{Introduction}
\label{introduction}

The photospheric lithium abundance in cool, low-mass stars probes their internal structure and mixing processes. The dominant $^7$Li isotope undergoes fusion at lower temperatures ($\sim 3\times 10^6$ K) than hydrogen, so ``Li-burning'' begins in the contracting pre main sequence (PMS) phase once central temperatures reach this ignition threshold \citep[e.g.,][]{Bildsten1997a}. In the lowest mass PMS stars that remain fully convective all the way to the zero age main sequence (ZAMS), mixing from centre to surface is efficient, and rapid photospheric Li depletion is predicted. For higher mass stars ($\geq 0.4\,M_\odot$), a growing radiative core provides a barrier to deep mixing; in models of this process, photospheric Li depletion is arrested, leaving a mass-dependent, non-zero Li abundance at the ZAMS (e.g., \citealp{Pinsonneault1997a}, \citealp{Baraffe1998a}, \citealp*{Siess2000a}, \citealp{Piau2002a}, \citealp{Dotter2008a}, \citealp*{Tognelli2011a}, \citealp{Baraffe2015a}).

Different models predict different Li-depletion rates as a function of mass (or effective temperature, $T_{\rm eff}$), depending on the adopted opacities, equations of state, treatment of convective mixing and the atmosphere bounding the star \citep*[e.g.,][]{Tognelli2012a}. However, all agree that the Li-depletion pattern should be a single-valued function of mass/$T_{\rm eff}$, without dispersion. Observations contradict this prediction. In the Pleiades at an age $\sim 120$ Myr \citep{Duncan1983a, Butler1987a, Soderblom1993a, Jeffries1999a}, and since confirmed in several other clusters at somewhat younger and older ages (e.g., \citealp{Randich1998a}, \citealp*{Jeffries1998a}, \citealp{Jeffries1999b}, \citealp{Randich2001a}, \citealp*{Balachandran2011a}), it has been established that ZAMS stars with $4000 \leq T_{\rm eff}/K \leq 5400$ have an intrinsic Li abundance dispersion of up to 2 orders of magnitude, inferred from a corresponding range of $\sim 300$\,m\AA\ in the equivalent width (EW$_{\rm Li}$) of the Li~{\sc i}~6708\AA\ resonance line, that is far greater than any observational uncertainties.

There are several proposed explanations for the Li dispersion. The photospheres of these young, magnetically active stars are affected by spots, plages and overlying chromospheres, such that EW$_{\rm Li}$ may not accurately reflect Li abundances when interpreted with simple, homogeneous, LTE models \citep*{Stuik1997a, King2000a}; although other studies suggest these effects are too small to fully explain the observed dispersion \citep{Barrado2001b, King2010a}. It has also become clear that the dispersion is strongly correlated with rotation rate - the fastest rotating stars at a given $T_{\rm eff}$ have larger EW$_{\rm Li}$ \citep{Barrado2016a, Bouvier2018a} and this leads to the idea that the structural effects of dynamo-generated interior magnetic fields or surface starspots might inhibit convection, inflate the star compared to standard model expectations and delay or slow the rate of Li depletion in fast rotators \citep[e.g.,][]{Somers2015a, Somers2017a, Jeffries2021a}, or that the convection criterion is modified in faster rotators leading to shallower convection \citep{Constantino2021a}. Others suggest that slower rotators undergo additional mixing and Li depletion after their radiative cores have developed, as a result of angular momentum loss and consequent core-envelope decoupling \citep{Bouvier2008a}, or that faster rotators have less effective convective overshooting into the core \citep{Baraffe2017a}.

The solution to this PMS Li-depletion problem is important in understanding the complex interiors of stars, not only in the PMS phase but on the main sequence too. Ongoing mixing is required to explain why the Sun and other cool stars are heavily Li-depleted; ZAMS Li abundances provide the approximate starting point for those processes. Understanding the dispersion is also important in: interpreting a similar Li abundance scatter in older stars in the field and in clusters, where their rotational history may be important; using lithium to identify young stars or as an age indicator; and interpreting Li abundances in the context, and as a probe, of Galactic chemical evolution \citep[e.g,][]{Guiglion2016a, Gutierrez2020a, Romano2021a, Jeffries2023a, Saad2024a}.

In this paper, we attempt to put observational constraints on the potential mechanisms that drive the Li dispersion by identifying the age at which the scatter appears and the timescale on which it develops on the PMS. Previous heterogeneous studies of individual clusters and associations have suggested that the rotation-dependent spread is already established at 45-60 Myr\footnote{See \protect\citet*{Jeffries2023b} and references therein on the ages of these clusters.} in the IC~2391/2602 clusters \citep{Randich2001a}, is probably present in NGC~2547 at an age of 35-40\,Myr \citep{Binks2022a} and at ages of 20-25 Myr in the Beta Pic association \citep{Messina2016a}, and possibly even as early as $\sim 5$ Myr in NGC~2264 \citep{Bouvier2016a}. Here, we extend these studies by providing an analysis of a homogeneous dataset of EW$_{\rm Li}$/$T_{\rm eff}$ measurements for thousands of stars in young clusters with well-determined relative ages and secure membership, which were observed as part of the {\it Gaia}-ESO spectroscopic survey (GES).

In \S\ref{sec:obs} the observational sample is described and \S\ref{measured_dispersion} explains how the dispersion in EW$_{\rm Li}$ is measured and examines its dependence on age, mass and rotation. In \S\ref{spot_model} a model based on a variable starspot coverage is investigated in an attempt to explain the EW$_{\rm Li}$ scatter and the necessary requirements of such a model are established. \S\ref{sec:discussion} discusses whether these requirements are likely to be met and also compares the measured Li dispersion with other possible explanations.

\section{The observational sample}
\label{sec:obs}

\subsection{The parent catalogue}

\label{parent_catalogue}

The principle source of data used in this paper are medium and high-resolution spectra of many thousands of stars towards open clusters at a range of ages, obtained as part of GES \citep[see][for details of the GES survey, instrumentation, target selection towards clusters and observations]{Gilmore2022a, Randich2022a, Bragaglia2022a}. A kinematic cluster membership study was performed by \cite{Jackson2022a} using the radial velocities and spectroscopic measurements of $T_{\rm eff}$ and gravity from GES, combined with proper motions from the {\it Gaia} DR3 catalogue. An important point is that neither colour, magnitude or chemical abundances (including Li) were used as membership criteria, leading to membership probabilities unbiased with respect to these properties.

A subset of these stars, with kinematic membership probabilities $>0.9$ (with an expected contamination rate of just 0.6 per cent),  $2900<T_{\rm eff}/K<6600$ and with probable giants excluded, were homogeneously analysed in \cite{Jeffries2023a} to obtain EW$_{\rm Li}$. The EW$_{\rm Li}$ measurements were based on a template subtraction technique, using presumptively Li-poor older field stars from the same GES observations. This method accounts (to first order) for any blending with other lines in the spectrum -- notably, an Fe~{\sc i} line at 6707.4\AA\ and various molecular features in the coolest stars. The templates were also appropriately broadened according to the estimated projected equatorial velocities ($v \sin i$) of the targets. EW$_{\rm Li}$ measurements with uncertainties were reported, even where the measurement is consistent with Li being completely depleted. 

Here, from the full list of 52 clusters reported in \cite{Jeffries2023a} a subset of 17 with age\,$<$500\,Myr were selected, which have a sufficient number of targets ($\ge 7$ after filtering) in the mass range equivalent to K-dwarfs at the ZAMS (see \S\ref{mass_ranges}), to provide a useful measure of dispersion. To this were added data from two large, well-studied non-GES ZAMS clusters -- M35 and the Pleiades: EW$_{\rm Li}$ values and uncertainties for the Pleiades were taken from \cite{Soderblom1993a} (corrected for blending), with $T_{\rm eff}$ calculated from the Gaia DR3 $G_{Bp}-G_{Rp}$ colour as described in \S4.3 of  \cite{Jeffries2023a}; data for M35 were taken from \cite{Jeffries2021a}. Targets with S/N $< 10$, $v\sin i >200$\, km\,s$^{-1}$ or EW$_{\rm Li}$ measurement uncertainty $>80$\,m\AA~were discarded at this stage and after additional filtering (see \S\ref{additional_filtering}) gave a data set of 2472 targets in 19 clusters (see Table \ref{target_data}).

The clusters used are listed in Table~\ref{cluster_numbers} along with their ages, which are the fiducial ``training" ages adopted and described in \cite{Jeffries2023a}. To briefly recap, these ages are a geometric mean (i.e. the mean log age) of: ages compiled from the literature by \cite{Jackson2022a}; the independent literature ages \citep[mostly from][]{Cantat-Gaudin2020a} listed in tables~3 and~4 of \cite{Randich2022a}; and the age determinations of \cite{Dias2021a} based on consistent isochrone fitting to {\it Gaia} photometry. The one exception is IC~4665, where the \cite{Jackson2022a} literature age was found to be based on an erroneous analysis and was revised upwards to 55\,Myr \citep[see][]{Jeffries2023b}.
The adopted ages for any individual cluster are of course subject to some uncertainty and the overall age scale is model-dependent. As demonstrated in \cite{Jeffries2023a}, the adopted cluster ages in the range 20-150\,Myr, that are most important in this paper, are quite consistent with ``lithium depletion boundary" ages, that use the almost model-independent luminosity of the faintest fully Li-depleted M-dwarf in a cluster as an age indicator \citep*[e.g.,][]{Burke2004a, Tognelli2015a}.

\subsection{Stellar Models}
\label{models}

To follow the evolution of Li abundance and to define relationships beween mass, radius, $T_{\rm eff}$ and age, grids from two stellar evolutionary models were used -- the {\sc spots} models \citep*{Somers2020a} and a version of the Pisa models \citep{Tognelli2011a, Tognelli2021a} modified to include the effects of starspot coverage \citep[see][and references therein]{Franciosini2022a}. Both provide grids of stellar evolution tracks and isochrones at varying levels of spot coverage but differ in their input physics, the list of output parameters reported and the lower limit of stellar mass considered. For these reasons, results from both models were used and compared where possible.

The {\sc spots} models provide a grid for stars with mass $\ge 0.1\,M_{\odot}$ with a fixed  ratio, $x_{\rm spot}=0.8$, between the spot temperature and the unspotted photospheric temperature, for six levels of fractional spot areal coverage, $0 \leq f_{\rm spot} \leq 0.85$. In most respects this is equivalent to a fractional photospheric flux blocking factor, $0 \leq \beta \leq 0.5$ \citep[see \S2 of][]{Somers2020a}. The Pisa models give a finer grid of stellar mass $\ge 0.08\,M_{\odot}$ at four levels of $\beta$, between 0 and 0.4. Results are available for three mixing length parameters; the solar-calibrated value of $\alpha_{\rm ML}=2.0$ were used here. Neither set of models allows the value of $\beta$ to vary with age.

\subsection{Mass ranges}
\label{mass_ranges}

In what follows, our focus is to understand how the EW$_{\rm Li}$ dispersion evolves for stars in three mass ranges corresponding to the following $T_{\rm eff}$ ranges at the ZAMS\footnote{Practically, the ZAMS was defined to be the isochrone at 120\,Myr. This simple definition is a good approximation for the stars considered in this work.}:  (1) early M-dwarfs ($3450 < T_{\rm eff}/{\rm K}<4000$); 
(2) K-dwarfs ($4080 < T_{\rm eff}/{\rm K}<5200$); (3) late G-dwarfs ($5300 < T_{\rm eff}/{\rm K} < 5710$) The two stellar models described in \S\ref{models} were used to define the equivalent mass ranges (see Table~\ref{ranges}) and to calculate the corresponding $T_{\rm eff}$ ranges at other ages.

The adopted $T_{\rm eff}$ limits at other ages depend, albeit weakly, on the assumed levels of spot coverage. \cite{Cao2022a}~ measured the spot coverage for ZAMS K- and M-dwarfs in the Pleiades. Using results from their table~1, we find an approximately lognormal distribution with a mean of $\log{\beta } = -1.066$ for all single stars. Analysis of younger clusters may suggest a higher average $\beta$. \cite{Cao2022b} used a prior of $f_{\rm spot}=0.34$ for the {\sc spots} model (equivalent to $\beta=0.2$) in their analysis of the $\sim 10$\,Myr $\lambda$~Ori cluster, based on fitting the slope of the mass/age relationship and eclipsing binary data; \cite{Binks2022a} and \cite{Franciosini2022a} found isochrones with $\beta \geq 0.2$ gave better simultaneous matches to young cluster data in colour-magnitude diagrams and the Li-$T_{\rm eff}$ plane. On these bases, we assumed $\beta=0.2$ for PMS stars aged $< 75$\,Myr and $\beta=0.1$ for older clusters. An increase of 0.1 in $\beta$ for the younger clusters decreases both upper and lower $T_{\rm eff}$ limits  by $\sim$100\,K for all the mass ranges considered.

Table \ref{cluster_numbers} shows in detail the mass and age-dependent $T_{\rm eff}$ limits for the K-dwarf range in each cluster. The {\sc spots} grid yields a mass range of $0.63< M/M_{\odot}<0.90$~compared to  $0.60<M/M_{\odot}<0.85$ for the Pisa grid, indicating small differences in the mass-$T_{\rm eff}$ relation between the two sets of models at 120 Myr, however there are just $\sim 10-20$\,K differences in the $T_{\rm eff}$ limits over the full range in ages considered.

\subsection{Additional filtering}
\label{additional_filtering}
\begin{figure*}
	\begin{minipage}[t]{0.98\textwidth}
            \includegraphics[width = 175mm]{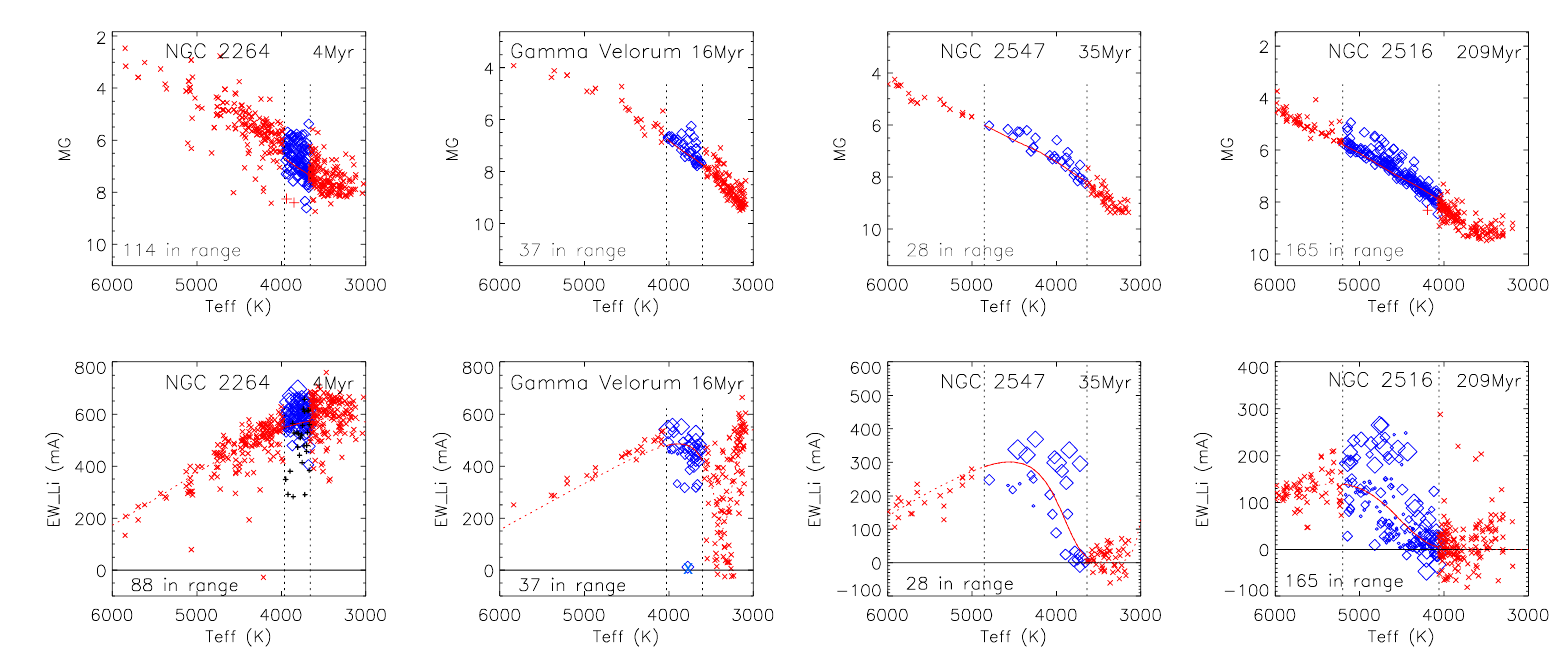}
        \end{minipage}
	\caption{Upper plots show $M_G$ versus $T_{\rm eff}$ for four relatively well populated clusters using data from Table \ref{target_data}. Vertical dashed lines indicate the $T_{\rm eff}$ limits corresponding to the K-dwarf mass range. Targets within these limits shown as blue diamonds are those remaining after filtering by  $M_G$, with the red line indicating the empirical isochrone for single stars  (see \S\ref{additional_filtering}). Lower plots show EW$_{\rm Li}$ versus $T_{\rm eff}$ for the same clusters and mass range. Blue diamonds indicate valid targets, with symbol size scaled in this case by $\log{v\sin{i}}$; the red line (dashed outside the considered mass range) shows the best-fitting {\sc eagles} EW$_{\rm Li}$ isochrone \citep{Jeffries2023a}, which is our baseline to assess the EW$_{\rm Li}$ dispersion. Black crosses show the targets rejected as being accreting sources with $\alpha_c>4$. The target shown as a crossed diamond in  Gamma Velorum was rejected as an extreme EW$_{\rm Li}$ outlier (see Table \ref{target_data}). Targets outside the mass range are as Table \ref{target_data} with no additional filtering.}  
	\label{fig1}	
\end{figure*}

Additional filtering was undertaken to remove a few possible contaminants that, although passing kinematic tests of cluster membership, were unlikely cluster members based on their parallax or photometry. Targets with parallax >3$\sigma$ from the cluster mean were rejected. Absolute $G$ magnitude ($M_G$) was calculated  using extinctions and distance moduli from \cite{Jackson2022a} for GES targets and from \cite{Jeffries2023a} for the Pleiades and M35, assuming $A_G=2.5E(B-V)$ \citep{Chen2019a}. An empirical $M_G$ versus $T_{\rm eff}$ isochrone (for probable single stars in the $T_{\rm eff}$ range corresponding to the subsample being considered -- see \S\ref{mass_ranges}) was defined for each cluster by offsetting a Pisa model isochrone, at the cluster age, to match the median absolute offset in $M_G$ for stars with an offset less than the median (i.e. likely single stars). Targets with an offset $>4\sigma$ below, or $>0.75$ mag $+4\sigma$ above, this empirical single-star isochrone were rejected, where $\sigma$ is the median absolute dispersion of probable single stars relative to the empirical isochrone (e.g., see Fig.~\ref{fig1} for this applied in the ZAMS K-dwarf mass range for several clusters). The photometric selection criteria exclude very few objects.

A further filter was applied to remove targets with evidence of significant accretion activity, since EW$_{\rm Li}$ could be underestimated due to a veiling continuum. Targets with very large, and presumably accretion-related, H$\alpha$ emission lines, quantified by a H$\alpha$ core index \citep{Damiani2014a} $\alpha _c \ge 4$, were excluded from the analysis of dispersion, as were two extreme EW$_{\rm Li}$ outliers: 08490442-4127148 in Ascc\,50, 08093642-4717442 in Gamma Velorum  (see Fig.~\ref{fig1}).

The initial selection described in \S\ref{parent_catalogue} yielded 2795 targets in 19 clusters of which 2649 have parallax data $<3\sigma$ from the cluster mean. Of these, 177 show a high level of  $\alpha_c \ge 4$, all but 4 of which are in the clusters aged $<10 $\,Myr. Table~\ref{target_data} lists all 2649 targets. The penultimate column flags targets assigned to the G-, K- and M-dwarf mass ranges. The final column is a flag set to 1 if targets in those mass ranges have $\alpha_c < 4$, or 0 otherwise. The flag is also set to zero for 8 targets which are excluded from the final analysis, 6  on the basis of being photometric outliers and 2 being the extreme EW$_{\rm Li}$ outliers.

\begin{table}
\caption{Basic data for stars in the 19 analysed clusters. The format and content of the Table is given here; the full table, with 2649 rows, is available in electronic format.}
\begin{tabular}{lll}
\hline
Col  & Label & (Units)/Description \\
\hline
1  & Cluster & Cluster identifier \\
2  & Target  & Name  \\
3  & IDGaia & {\it Gaia} DR3 ID \\
4  & SNR & Signal to noise ratio of source spectrum \\
5  & vsini & ( km\,s$^{-1}$) projected equatorial velocity$^*$ \\
6  & Teff & (K) Effective temperature\\
7  & plx & (mas) parallax (from Gaia DR3 catalogue) \\
8  & mem & Membership probability \citep{Jackson2022a}\\
9  & Gmag & (mag) $G$ mag. (from Gaia DR3 catalogue) \\
10 & Bp\_ Rp & (mag) $G_{Bp}-G_{Rp}$ (from Gaia DR3 catalogue) \\
11 & EWLi & (m\AA) Lithium equivalent width, EW$_{\rm Li}$ \\
12 & eEWLi & (m\AA) measurement uncertainty  \\
13 & Period & (d) Expectation value calculated from $v\sin{i}^{**}$\\
14 & Alc   & H-alpha core index \citep{Damiani2014a}\\
15 & Range & Letter indicating mass range using {\sc spots} model\\
16 & Filter  & Flag indicating a filtered target (see \S\ref{additional_filtering})\\
\hline
\multicolumn{3}{l}{* Reduced $v\sin i$ from \cite{Randich2022a}.}\\
\multicolumn{3}{l}{** Measured periods shown for the Pleiades and M35.}\\
\end{tabular}
\label{target_data}
\end{table}

\begin{table}
\caption{Cluster names and ages from \protect\cite{Jeffries2023a}, with calculated temperature ranges and resulting target numbers for the K-dwarf mass range.}
\begin{tabular}{l@{\hspace{1mm}}l@{\hspace{0mm}}lll@{\hspace{1mm}}lll}
\hline
Cluster & Fiducial & \multicolumn{3}{c}{{\sc spots} Model} &  \multicolumn{3}{c}{Pisa Model} \\
 & Age  & \multicolumn{3}{l}{0.63$<$ $M/M_{\odot}$ $<$0.90} & \multicolumn{3}{l}{0.60$<$ $M/M_{\odot}$ $<$0.85} \\
        & (Myr)  & $T_{\rm eff}^{\rm min}$ & $T_{\rm 
 eff}^{\rm max}$ & No. & $T_{\rm eff}^{\rm min}$ & $T_{\rm eff}^{\rm max}$ & No.\\
\hline
NGC\,2264	& 4.5 &	3655	&	3961	&	88	&	3648	&	3953	&	94	\\
NGC\,2244	& 6.0 &	3728	&	4044	&	21	&	3727	&	4050	&	21	\\
Ascc\,50	& 8.1 &	3711	&	4023	&	45	&	3708	&	4010	&	45	\\
Lambda Ori	&	8.7 & 3619	&	3951	&	26	&	3611	&	3952	&	29	\\
Collinder\,197 & 11.9 & 3643	&	3952	&	15	&	3634	&	3947	&	15	\\
25\,Ori	    &	15 & 3601	&	4018	&	31	&	3598	&	4034	&	34	\\
Gamma Vel	&	16 & 3600	&	4027	&	36	&	3598	&	4044	&	36	\\
NGC\,2232	&	27 & 3615	&	4534	&	15	&	3619	&	4546	&	14	\\
NGC\,2547	&	35 & 3638	&	4856	&	28	&	3643	&	4895	&	28	\\
NGC\,2451b	&	38 & 3697	&	5255	&	19	&	3703	&	5290	&	20	\\
NGC2451a	&	50 & 4002	&	5143	&	9	&	4035	&	5160	&	9	\\
IC\,4665	&	55 & 3814	&	5161	&	7	&	3823	&	5174	&	7	\\
NGC\,6405	&	63 & 4134	&	5204	&	19	&	4157	&	5206	&	19	\\
Blanco\,1	&	110 & 3985	&	5143	&	16	&	4012	&	5160	&	14	\\
Pleiades	&	120 & 4162	&	5201	&	48	&	4194	&	5202	&	45	\\
M35	        &	140 & 4098 &	5200	&	129	&	4117	&	5200	&	129	\\
NGC\,6709	& 166 & 4060	&	5199	&	12	&	4067	&	5199	&	12	\\
NGC\,2516	& 209 &	4060	&	5199	&	165	&	4067	&	5199	&	164	\\
NGC\,3532	& 367 &	4040	&	5203	&	114	&	4054	&	5198	&	113	\\
\hline
\end{tabular}
\label{cluster_numbers}
\end{table}

\begin{table}
\caption{Mass ranges and corresponding $T_{\rm eff}$ ranges at the ZAMS.}
\begin{tabular}{lccc}
\hline
Range		&	$T_{\rm eff}$ range	& \multicolumn{2}{c}{Mass range ($M_\odot$)} \\
Name		&	 at ZAMS	&	{\sc spots} models &	 Pisa models\\
\hline
Early M	&	3450…4000\,K	&	0.41…0.61	&	0.37…0.57\\
K		&	4080…5200\,K	&	0.63…0.90	&	0.60…0.85\\
Late G	&	5300…5710\,K	&	0.93…1.05	&	0.87…0.98\\
\hline
\end{tabular}
\label{ranges}
\end{table}

\begin{figure*}
	\begin{minipage}[t]{0.98\textwidth}
            \includegraphics[width = 175mm]{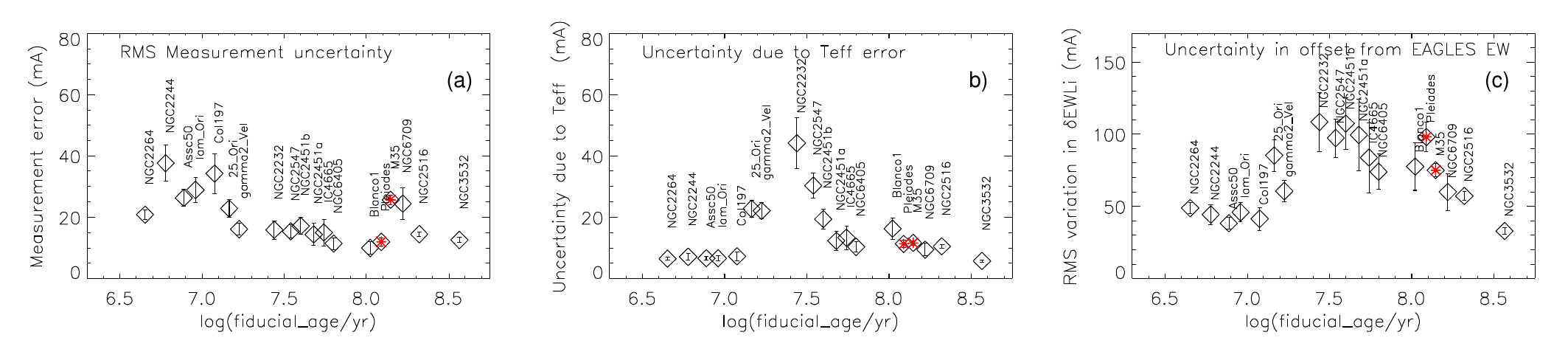}
        \end{minipage}
	\caption{RMS variation in EW$_{\rm Li}$ for individual clusters, as a function of fiducial age (column 2 in Table~\ref{cluster_numbers}), for targets in the ZAMS K-dwarf mass range (from the {\sc spots} models, see Table~\ref{cluster_numbers}). Plot (a) shows $\sigma_{\rm meas}$, the RMS measurement uncertainty in EW$_{\rm Li}$.  Plot (b) shows $\sigma_{\rm Teff}$, the estimated RMS uncertainty in EW$_{\rm Li}$ contributed by measurement errors in $T_{\rm eff}$ (see \S\ref{measured_dispersion}). Plot (c) shows $\sigma_{\rm eff}$, the RMS variation in the offset in EW$_{\rm Li}$ of individual targets from the best-fitting EAGLES isochrone for each cluster \citep{Jeffries2023a}. The plotted ages are offset slightly to clearly identify each cluster. Red symbols indicate the 2 non-GES clusters.}
	\label{fig2}	
\end{figure*}

\begin{figure*}
	\begin{minipage}[t]{0.98\textwidth}
            \includegraphics[width = 170mm]{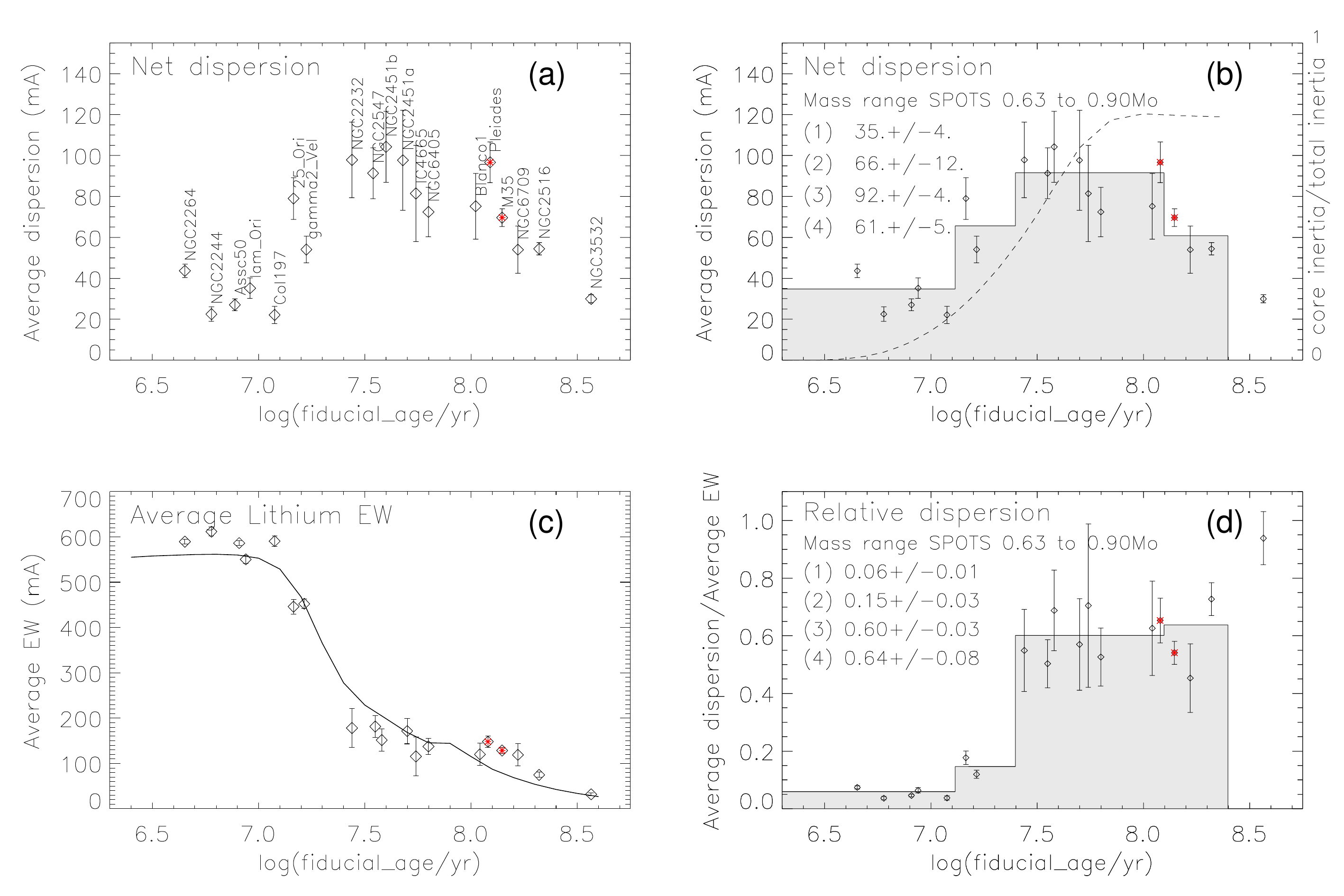}
        \end{minipage}
	\caption{Intrinsic dispersion of stars in the ZAMS K-dwarf mass range (according to the {\sc spots} models) as a function of age. Plot (a) shows $\Delta$EW$_{\rm Li}$ (see Eqn.~\ref{Eqn1}) for individual clusters as a function of $\log$~age, with small offsets to identify each cluster. Plot (b) shows the same data without age offsets, with a histogram and text on the plot showing the weighted mean $\Delta$EW$_{\rm Li}$ (in m\AA) in four age bands (see \S\ref{K-dwarf_dispersion}). The dashed line illustrates the growth of the radiative core (from the {\sc spots} models), parameterised by the fraction of the stellar moment of inertia that is within the radiative core plotted on the right-hand y-axis. Plot (c) shows the average EW$_{\rm Li}$ for individual clusters, $\langle{\rm EW}_{\rm Li}\rangle$, with a curve plotting the {\sc eagles}-predicted mean value over the corresponding $T_{\rm eff}$ range. Plot (d) shows a normalised dispersion, $\Delta$EW$_{\rm Li}/\langle{\rm EW}_{\rm Li}\rangle$ for individual clusters, with a histogram and text showing weighted average values in four age bands.}
	\label{fig3}	
\end{figure*}

\begin{figure}
    \includegraphics[width = 85mm]{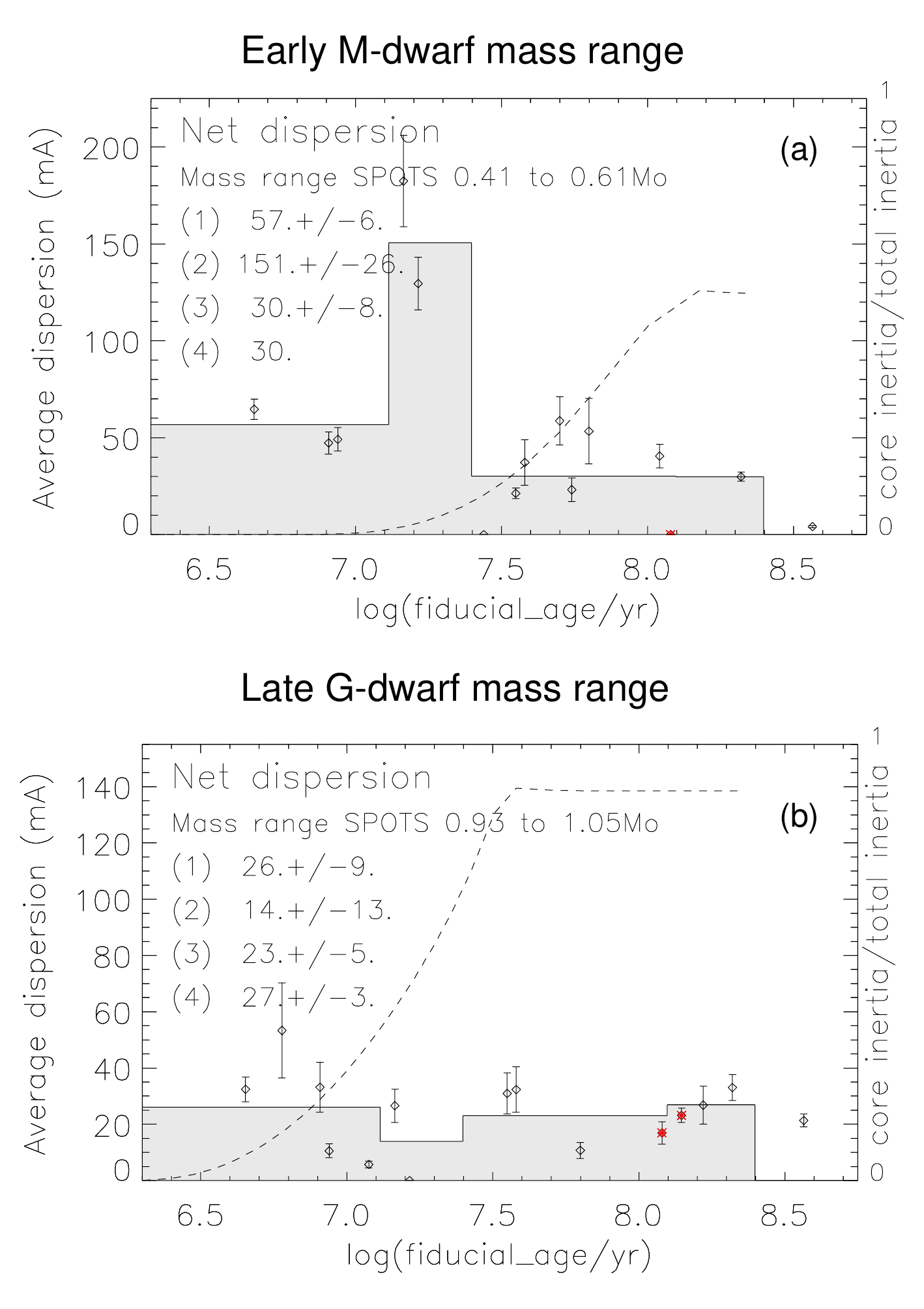}
	\caption{Empirical dispersion as a function of age and spectral type for targets in the early M and late G dwarf mass ranges (see Table \ref{ranges}). Plots show the RMS level of dispersion, $\Delta$EW$_{\rm Li}$ for clusters with 5 or more targets in the specified mass range together with a histogram and text on the plot showing the weighted mean dispersion in four age bands (see \S \ref {spectral type}). The dashed lines with a corresponding scale on the right-hand y-axis show the fraction of the stellar moment of inertia contributed by a radiative core.}
\label{fig4}	
\end{figure}

\section{The Measured Dispersion}
\label{measured_dispersion}

We begin the analysis by assessing the dispersion of Li and how it evolves in an observational sense, using the evolution of measured dispersion of EW$_{\rm Li}$ in each cluster about an age-dependent mean relationship with $T_{\rm eff}$, corrected for the contribution to dispersion from observational uncertainties. We explicitly assume that each cluster is coeval and that none of the observed dispersion relates to any age dispersion within a cluster. Of course, to relate mean level and dispersion in EW$_{\rm Li}$ to changes in Li {\it abundance} requires both stellar evolutionary and stellar atmosphere models. These considerations are deferred to \S\ref{spot_model} in order to focus here on the observational data.

Figure~\ref{fig1} shows examples of target selection in the ZAMS K-dwarf mass range for four well-populated clusters in the $M_G$ vs $T_{\rm eff}$ and EW$_{\rm Li}$ vs $T_{\rm eff}$ diagrams. Vertical dashed lines indicate the temperature limits based on the {\sc spots} model (see Table~\ref{cluster_numbers}). Targets within the mass range (shown in blue) are those remaining after filtering as described in \S\ref{additional_filtering}. In the $M_G$ vs $T_{\rm eff}$ plots, the red lines between the dashed $T_{\rm eff}$ limits indicate the empirical single star isochrone that was used. In the EW$_{\rm Li}$ vs $T_{\rm eff}$ plots, the symbol size of the selected targets is scaled as $\log{v\sin i}$ and red lines show best-fitting empirical isochrones produced by the {\sc eagles} code to the cluster data \citep[see][]{Jeffries2023a}, which are used as the baseline from which to estimate the EW$_{\rm Li}$ dispersion.
In NGC~2264, targets that meet the kinematic criteria for cluster membership but have $\alpha _c>4$, indicative of significant accretion activity, are marked with black crosses.

For each cluster, the intrinsic EW$_{\rm Li}$ dispersion, $\Delta$EW$_{\rm Li}$, was estimated as the RMS variation of the offset in EW$_{\rm Li}$ of individual targets relative to the {\sc eagles} isochrone ($\sigma_{\rm off}$), after subtracting (in quadrature) the estimated uncertainties in EW$_{\rm Li}$ contributed by measurement errors in EW$_{\rm Li}$ ($\sigma_{\rm meas}$) and $T_{\rm eff}$ ($\sigma_{\rm Teff}$) as follows: 
\begin{eqnarray}
{\rm if\,\,} \sigma_{\rm off}^2 & > & (\sigma_{\rm  meas}^2 + \sigma_{\rm Teff}^2)
\label{Eqn1} \\\nonumber 
\Delta {\rm EW}_{\rm Li} & = & \sqrt{\sigma_{\rm off}^2 - \sigma_{\rm meas}^2 - \sigma_{\rm Teff}^2}\ , \\\nonumber
{\rm else\ \ \ \ }  \\ \nonumber
\Delta {\rm EW}_{\rm Li} & = & 0\ .
\end{eqnarray}
 $\sigma_{\rm meas}$ is the RMS value of the EW$_{\rm Li}$ errors in Table~\ref{target_data}.  $\sigma_{\rm Teff}$~is estimated as the RMS change in the {\sc eagles}-predicted EW$_{\rm Li}$ produced by the $\pm 80$\,K RMS measurement error in $T_{\rm eff}$ for the GES targets. At this stage this merely accounts for the role of $T_{\rm eff}$ measurement uncertainties in inflating $\sigma_{\rm off}$ if stars followed the mean {\sc eagles} isochrone and makes no attempt to model or interpret any scatter in terms of spots or activity (see \S\ref{spot_model}).
 Figure~\ref{fig2} shows $\sigma_{\rm meas}$, $\sigma_{\rm Teff}$ and $\sigma_{\rm off}$ in each cluster, for targets in the ZAMS K-dwarf mass range in Table~\ref{cluster_numbers}.

\subsection{Dispersion in the ZAMS K-dwarf mass range}
\label{K-dwarf_dispersion}
Figure~\ref{fig3}a shows $\Delta$EW$_{\rm Li}$, calculated using Eqn.~\ref{Eqn1}, for targets in the ZAMS K-dwarf mass range, using $T_{\rm eff}$ limits from the {\sc spots} model. The plot has some small offsets in age to show the individual clusters and their names. Figure~\ref{fig3}b shows the same $\Delta$EW$_{\rm Li}$, without the age offsets, along with a histogram giving the weighted average dispersion in four broad age bands. Figure~\ref{fig3}c shows the mean level of Li equivalent width, $\langle{\rm EW}_{\rm Li}\rangle$, and the mean EW$_{\rm Li}$ predicted by {\sc eagles} as a function of age for the corresponding $T_{\rm eff}$~range. Figure~\ref{fig3}d shows a normalised dispersion, $\Delta$EW$_{\rm Li}/\langle{\rm EW}_{\rm Li}\rangle$. Given the pseudo-linear relationship between logarithmic Li abundance and $\log\,$EW$_{\rm Li}$ for the saturated Li~I line \citep[e.g.,][and discussed further in \S\ref{spot_model}]{Franciosini2022a}, then $\Delta$EW$_{\rm Li}/\langle{\rm EW}_{\rm Li}\rangle$ is roughly proportional to the dispersion in logarithmic Li abundance. The main results are:
\begin{itemize}
\item Clusters in the youngest age band, $\le 13$\,Myr, show little evidence of Li depletion. Whilst there is some inferred intrinsic dispersion, $\Delta $EW$_{\rm Li} \simeq 35$\,m\AA, it is only 6 per cent of $\langle{\rm EW}_{\rm Li}\rangle$. This level of dispersion could be due to a modest underestimate of the measurement uncertainties in EW$_{\rm Li}$ (see Fig.~\ref{fig2}) or the influence of accretion-related veiling remaining for some stars.
It could also be that these youngest clusters have some age spread that is a significant fraction of their mean age. This might introduce some dispersion if depletion has begun in the oldest stars of any age distribution, but Figure~\ref{fig3}c suggest that such stars would need to be older than $\sim 15$\,Myr. Age spreads $\leq 10$\,Myr \citep[a generous upper limit, see][]{Palla1999a, Hartmann2001a, Jeffries2011b} will not greatly inflate the dispersion in older clusters because $|d\,\langle{\rm EW}_{\rm Li}\rangle/dt|$ is not steep enough.

\item Clusters in the second age band, $13 < {\rm age/Myr} < 25$, show a modest reduction in $\langle{\rm EW}_{\rm Li}\rangle$, coupled with a doubling of both their absolute and normalised intrinsic dispersion. 

\item The third age band, $25 < {\rm age/Myr} < 125$, shows a further 40 per cent increase in $\Delta $EW$_{\rm Li}$ together with a marked reduction in $\langle {\rm EW}_{\rm Li}\rangle$, resulting in a sharp increase in the normalised dispersion.

\item The oldest age band, $125 < {\rm age/Myr} < 250$, shows a decreasing $\Delta $EW$_{\rm Li}$ and $\langle {\rm EW}_{\rm Li}\rangle$ leading to no significant change in normalised dispersion. At these ages and older some of the stars have EW$_{\rm Li}$ consistent with zero (i.e. they have depleted Li to undetectable levels) and the increasing fraction of such stars with age will also suppress the absolute dispersion. This becomes more obvious in the M dwarfs, presumably because of their faster Li depletion timescales (see \S\ref{spectral type}).
\end{itemize}
As a check on model-dependence, Fig.~\ref{figA1} is the equivalent of Fig.~\ref{fig3} for the K-dwarf mass range evaluated using  $T_{\rm eff}$ limits from the Pisa model (Table~\ref{cluster_numbers}). The results are very similar to those obtained using the {\sc spots} $T_{\rm eff}$ limits, with both the dispersion and normalised dispersion in the various age bands agreeing within their uncertainties.

\subsection{Variation of dispersion with mass}
\label{spectral type}
Figure~\ref{fig4}~shows $\Delta$EW$_{\rm Li}$~as a function of age for the stars that would be early M or late G-dwarfs at the ZAMS, with mass ranges defined in Table~\ref{ranges}, using corresponding $T_{\rm eff}$ limits from the {\sc spots} model. There are fewer sample stars in these lower and higher mass ranges, and clusters with $<5$ targets within the relevant $T_{\rm eff}$ range are omitted from the plots. For completeness, plots showing $\langle$EW$_{\rm Li}\rangle$ and $\Delta$EW$_{\rm Li}/\langle$EW$_{\rm Li}\rangle$ for the early M-dwarf and late G-dwarf mass ranges are also shown in Fig.~\ref{figA2}.

There is a clear mass-dependence to the pattern of Li depletion and growth of dispersion. 
The late G-dwarfs (at the ZAMS) show consistently low levels of absolute and normalised dispersion in all age bands, despite the reduction in $\langle$EW$_{\rm Li}\rangle$ becoming significant at $>20$ Myr. The mechanism producing dispersion in EW$_{\rm Li}$~during the period of Li-burning in ZAMS K-dwarfs appears to have much less effect on higher mass (ZAMS G-dwarf) stars.

In contrast, the early M-dwarfs (at the ZAMS) show a small $\Delta$EW$_{\rm Li}$~in the youngest age band, that triples for $13 < {\rm age/Myr} < 25$ and falls back to a low level at ages $>25$ Myr. The peak visible in Fig.~\ref{fig4}a is probably a consequence of the sharp transition between ages at which Li is not depleted and only slightly older ages at which Li is completely destroyed (see \S\ref{spot_model}). At both these extremes, we expect no intrinsic dispersion. On the other hand, in the intermediate regime, that corresponds to the small age range where Li is partially depleted, any mechanism capable of producing a dispersion has the possibility to manifest its effects. The fact that we see any dispersion for M-dwarfs in some of the older clusters may indicate a small underestimate of the observational uncertainties but also the presence of a few odd, outlying objects that still possess a small amount of Li at older ages -- for example see the objects with $T_{\rm eff}<4000$\,K and EW$_{\rm Li} \sim 200$\,m\AA\ in NGC~2516 in Fig.~\ref{fig1}.

\begin{figure}
        \includegraphics[width = 85mm]{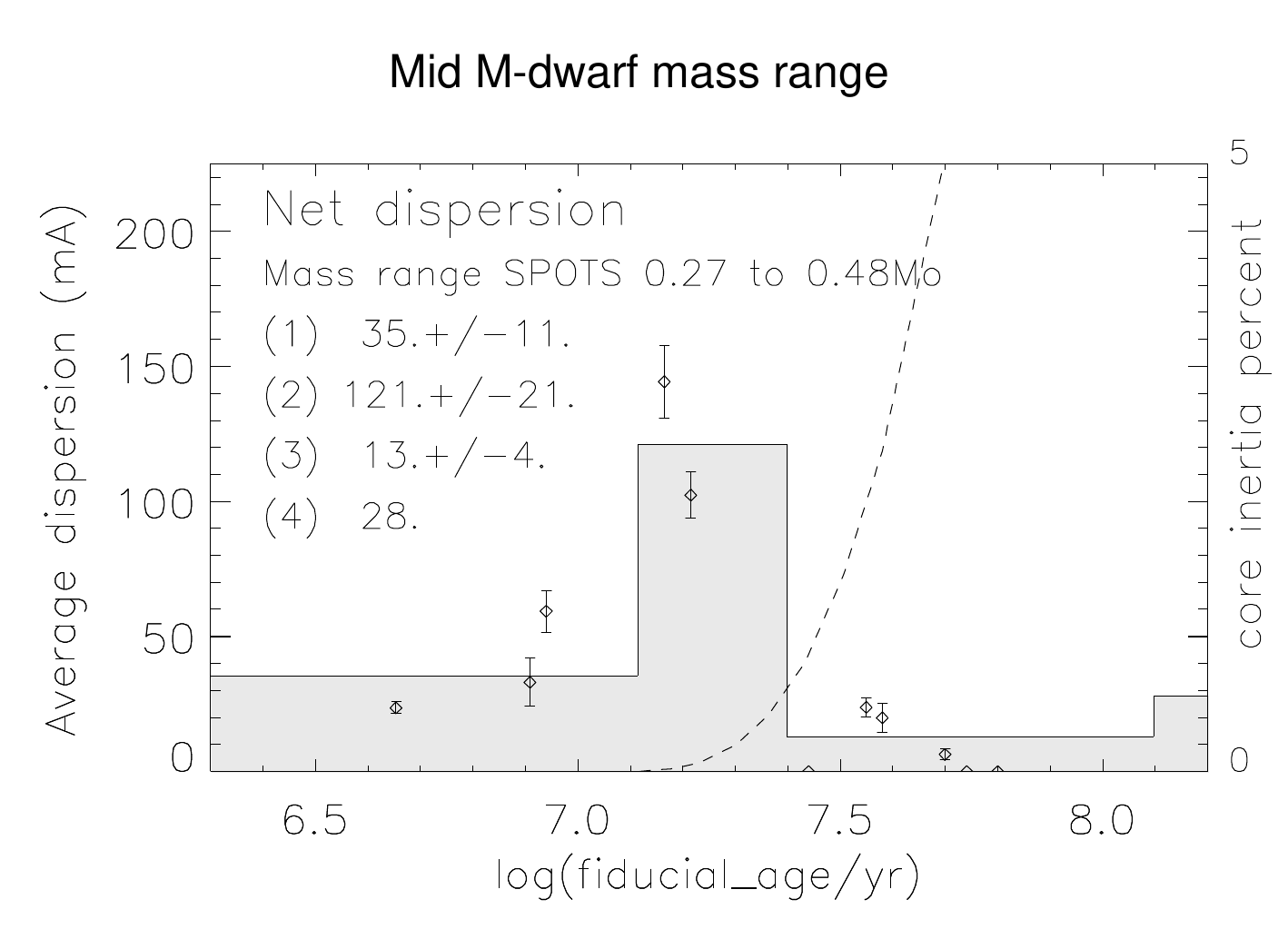}
  	\caption{Empirical dispersion, $\Delta$EW$_{\rm Li}$ versus log~age for the mid M-dwarf mass range (see \S \ref{radiative core}). Points show results for individual clusters with 5 or more targets in the mass range. The histogram and text on the plot shows the weighted mean dispersion in four age bands. The dashed line with a corresponding scale on the right-hand y-axis shows the percentage of the stellar moment inertia contributed by a radiative core between 0 and 5 per cent.}
       \label{fig9}	
\end{figure}

\begin{figure}
   \includegraphics[width = 85mm]{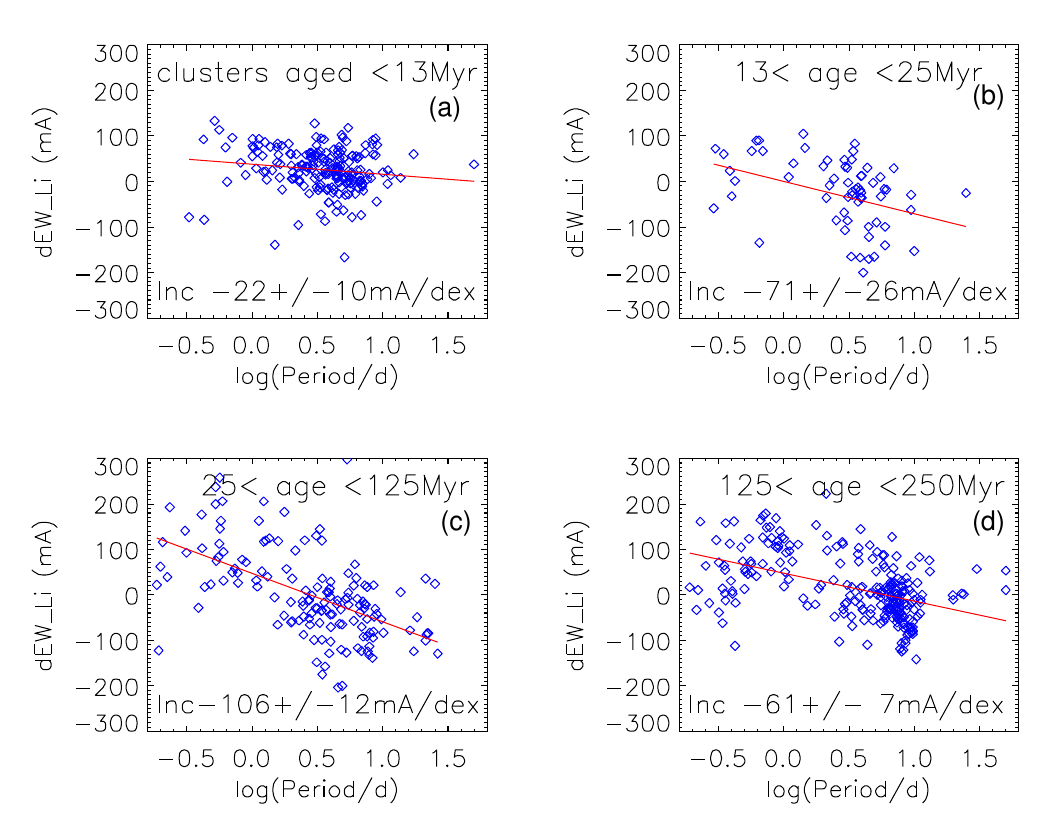}
   \caption{The offset of measured EW$_{\rm Li}$ for targets in the ZAMS K-dwarf mass range (from the {\sc spots} models) with respect to the {\sc eagles} EW$_{\rm Li}$ isochrone, at the $T_{\rm eff}$ and age of the target, versus rotation period estimated from  $v\sin{i}$ (see \S\ref{li_rotation}). Red lines show a linear least-squares fit; text on the plots gives the best fit slope and uncertainty.}
	\label{fig6}	
\end{figure}

\subsection{Influence of the stellar radiative core}
\label{radiative core}

The dashed lines in Figs.~\ref{fig3}b and~\ref{fig4} illustrate the growth of the radiative core, plotted as the fraction of the stellar moment of inertia that is contained within the radiative core (a parameter available from the {\sc spots} models).
For the ZAMS K-dwarfs the intrinsic dispersion in EW$_{\rm Li}$ develops in coincidence with the onset of Li depletion and possibly as the radiative core is formed. For the lower mass ZAMS early M-dwarfs it appears that Li depletion and dispersion is already well-established {\it before} the radiative core develops. To further explore this important point, we defined a sample of even lower-mass ZAMS mid M-dwarfs ($3250 < T_{\rm eff}/{\rm K} < 3600$ at 120\,Myr) using {\sc spots} model $T_{\rm eff}$ limits. These stars are expected to remain fully convective for longer and well past the epoch when Li is being depleted. The results comparing the development of $\Delta$EW$_{\rm Li}$ with that of the radiative core are shown in Fig.~\ref{fig9} (note that the right-hand y-axis upper limit is now just 5 per cent). This demonstrates that during the brief period between the onset of Li depletion at $\sim 10$\,Myr and almost total depletion at $\sim 20$ Myr, a significant EW$_{\rm Li}$ dispersion grows whilst the stars are predicted to be almost entirely convective.

\subsection{The Li-rotation correlation}
\label{li_rotation}

From previously published work (see references in \S\ref{introduction}) we know that some of the EW$_{\rm Li}$ dispersion is likely to be correlated with rotation. To investigate this, we grouped our sample stars into the same four age bins used in Fig.~\ref{fig3}b,d and plotted the offset of their EW$_{\rm Li}$ with respect to the {\sc eagles}-predicted EW$_{\rm Li}$, as a function of their estimated rotation period, $P_{\rm rot}$. For most of our sample (bar the Pleiades and M35, where measured rotation periods are those vetted and adopted by \citet{Bouvier2018a} and \citet{Jeffries2021a} respectively)  we have only $v \sin i$, so $P_{\rm rot}$ was estimated as
\begin{equation}
    P_{\rm rot} =  39.8 \left(\frac{v \sin i}{{\rm km\,s}^{-1}}\right)^{-1} \left(\frac{R}{R_\odot}\right)\ {\rm days}\ ,
\label{Prot}
\end{equation}
where the radius $R$ is estimated from the $T_{\rm eff}$ and age according to the {\sc spots} models and assuming the spin axes are randomly oriented \citep[e.g.,][]{Jackson2010a} so that $\langle \sin i \rangle = \pi/4$. The calculated (or actual) rotation periods are given in Table~\ref{target_data}.

Figure~\ref{fig6} shows the results for the ZAMS K-dwarf mass range (using the {\sc spots} models).   From this, a linear least squares fit is used to quantify the correlation that is apparent in all 4 age bands. The slope is shallow and of marginal significance at the youngest ages. If real, this suggests the small dispersion noted in \S\ref{K-dwarf_dispersion} could be due to some physical effect rather than underestimated uncertainties. The slope becomes steeper and more significant at 13--25\,Myr, which is when the majority of Li-burning is predicted to take place in these stars (see \S\ref{model_abundances}), steeper still and highly significant at 25--125\,Myr,  then a little shallower but still highly significant in the older ZAMS clusters.

We analysed the ZAMS early M-dwarf dispersion in a similar way but found no evidence for any significant Li-rotation correlation at ages $<25$ Myr, but with larger error bars on the slopes (see Fig:~\ref{figA3}). The lack of any correlation for ages $>25$ Myr is expected in these stars since almost all EW$_{\rm Li}$ values are consistent with zero.

\section{A Starspot Model for the Lithium Dispersion}
\label{spot_model}

Thus far the Li dispersion has been discussed in terms of the strength and dispersion of EW$_{\rm Li}$. This has the advantage of being observationally defined, but to connect it with the astrophysics of Li depletion we need to consider stellar models that predict the Li abundance evolution and fold these through a stellar atmosphere and line formation model (a curve of growth) to make predictions of EW$_{\rm Li}$ and its dispersion that can be compared with the observations.

A candidate explanation for the developing Li dispersion is that it is caused by a rotation-dependent dynamo and variable coverage of dark, magnetic starspots \citep{Somers2015a, Somers2015b, Somers2017a, Jeffries2021a}. The presence of spots changes the interior structure of the star; at a given age and mass, the radius is larger and the central temperature lower and this leads to a delayed onset and slower rate of Li depletion \citep{Somers2020a, Tognelli2021a}. The presence of a fractional coverage of cooler spots also changes the expected EW$_{\rm Li}$ for a given Li abundance because the curve of growth has a temperature dependence. Models that attempt to include the effects of spots have already been shown to provide better simultaneous fits to the colour-magnitude diagrams and Li-depletion patterns of young clusters \citep{Jeffries2017a, Binks2022a, Franciosini2022a}.  It is worth emphasising, that it is the dispersion in EW$_{\rm Li}$ that we are attempting to interpret, which is likely caused by a combination of a genuine Li abundance dispersion together with the effects of magnetic activity on the connection between abundance and EW$_{\rm Li}$. That is why the comparisons must be done in the observational space.

In this section, we investigate what spot coverage properties could lead to the observed magnitude, time-and rotation-dependence of the Li dispersion found in \S\ref{measured_dispersion}. In summary we:
\begin{itemize}
\item Use tabulated data from the {\sc spots} \citep{Somers2020a} and Pisa \citep{Tognelli2021a} models to predict values of Li abundance for a given age, $T_{\rm eff}$ and spot coverage. 

\item Convert the predicted Li abundances to model Li~6708\AA\ equivalent widths, EW$_{\rm m}$, using curves of growth, accounting for NLTE effects.

\item Set up a model with a hypothesised spot filling factor distribution and use this to predict a model EW$_{\rm m}$ dispersion,  $\Delta {\rm EW}_{\rm m}$, as a function of age, that can be compared in observational space with the results from \S\ref{measured_dispersion}.
\end{itemize}

\subsection{Model values of Lithium abundance}

\label{model_abundances}

\begin{figure}
   \includegraphics[width = 85mm]{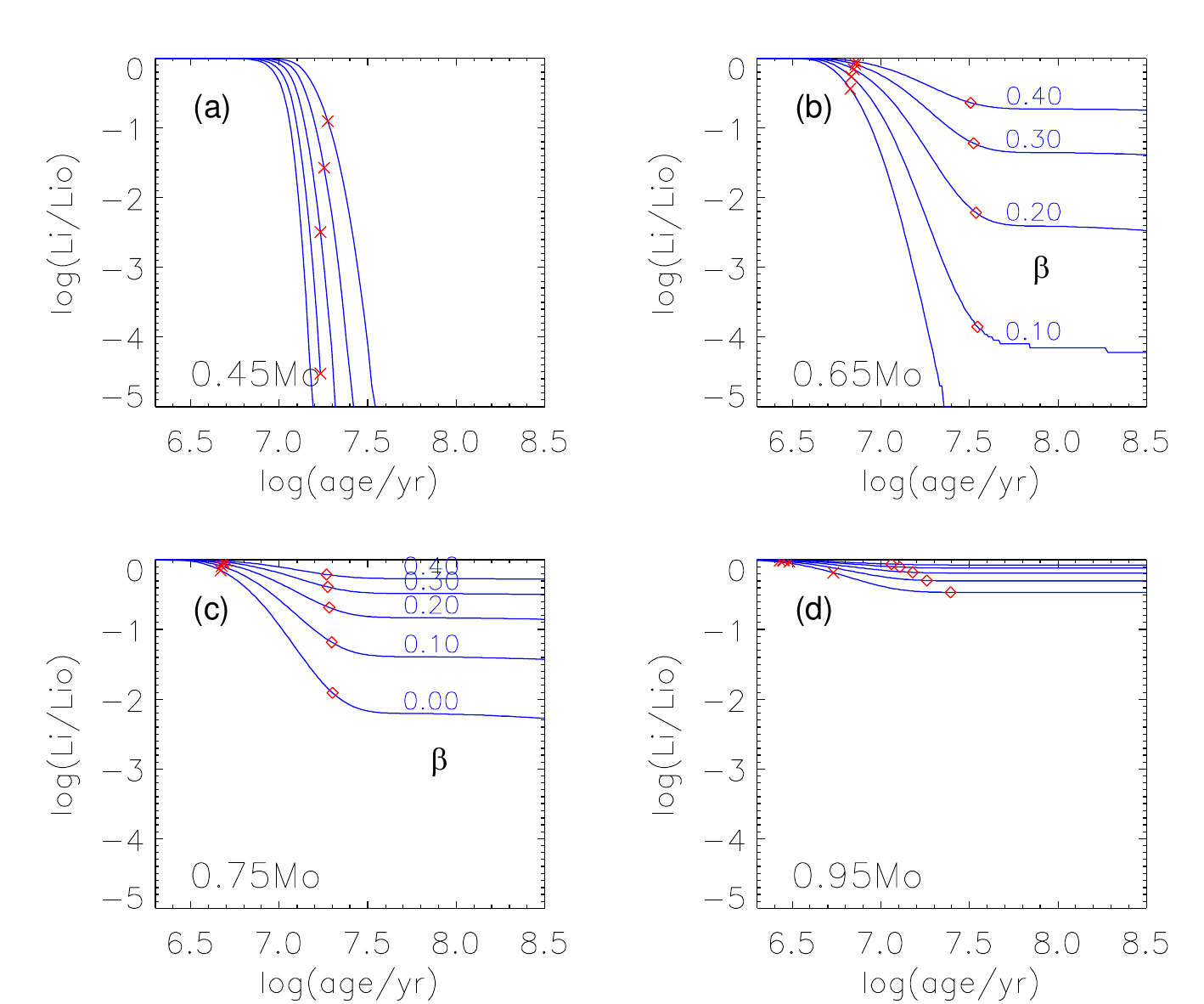}
   \caption{Evolutionary tracks of $\log_{10}$(Li/Li$_0$) from the {\sc spots} models for different levels of spot coverage (labelled with the flux blocking fraction $\beta$) and at 4 different masses. Red crosses and diamonds on each curve mark when the core inertia has reached 5 per cent and (for the three higher masses) 95 per cent of its ZAMS (120\,Myr) value.}
   \label{fig5}	
\end{figure}

Figure~\ref{fig5} shows evolutionary tracks of $\log_{10}$(Li/Li$_0$) from the {\sc spots} models at four different masses and for varying levels of flux blocking fraction $\beta$. Here, Li/Li$_0$ represents the ratio of the Li abundance to its initial value - assumed in what follows to be $A$(Li)$ = 3.30$ on the usual logarithmic scale with respect to $A$(H)$=12$. This value is consistent with the largest values of EW$_{\rm Li}$ in the youngest clusters (see \S\ref{model_ew}) and a compilation of $A$(Li) estimates in the F-stars of young GES clusters, that are not expected to have undergone any depletion, including several (NGC\,2264, IC\,4665, IC\,2602, NGC\,2547, NGC\,2516) used in this paper \citep{Randich2020a}. Evolutionary tracks from the Pisa models are quantitatively very similar.

\begin{itemize}
\item Figure~\ref{fig5}a shows tracks for a mass representative of the ZAMS early M-dwarf mass range. Whilst the evolution of Li/Li$_0$ is predicted to be strongly $\beta$-dependent, with a high $\beta$ value preserving Li for longer, Li is essentially gone by $\leq 30$\,Myr for all values of $\beta$ and well before stars reach the ZAMS. Li is already depleted by $\geq 1$ dex before the radiative core has grown to a significant size even for $\beta=0.4$ (see \S\ref{radiative core}). 

\item Figures~\ref{fig5}b,c are representative of the ZAMS K-dwarf mass range. Again, the results vary strongly with $\beta$ but in this case large, $\beta$-dependent differences in Li/Li$_0$ begin after $5-10$ Myr, as soon as Li-depletion starts and roughly coincident with the development of the radiative core, but these differences persist even after the ZAMS is reached and the Li abundance stabilises. This is because the stellar structure evolves very slowly on the main sequence, the convection zone base has become too cool to burn Li and the models do not include any additional, non-convective mixing.

\item At higher masses, representative of the  ZAMS late G-dwarf mass range (Fig.~\ref{fig5}d), a radiative core forms and the convection zone base falls below the Li-burning temperature earlier. As a consequence, photospheric Li depletion is small ($\leq 0.5$ dex), irrespective of $\beta$. Stars in this mass range could have only a small dispersion in Li abundance attributable to variations in spot coverage.  

Both sets of evolutionary models assume a solar composition -- $Z=0.0165$ and the \cite{Grevesse1998a} abundance mixture for the {\sc spots} models and $Z=0.0142$ and the \cite{Asplund2009a} abundance mixture for the Pisa models. Whilst there is little difference between the Li depletion predictions of these models, Li depletion is predicted to be quite composition-dependent with more PMS depletion at higher metallicity \citep[e.g.,][]{Piau2002a}. This should not contribute to the measured dispersion within clusters, since the stars within clusters should be reasonably chemically homogeneous \citep[apart from Li, e.g., $\sigma$(\protect{[Fe/H]})$<0.03$ dex in the Pleiades and Hyades][]{Liu2016a, Grilo2024a}.
Inter-cluster composition differences could contribute to the overall dispersion when stars are averaged in bins of age, however we note that this effect is unimportant because: the deviation that each star contributes to this dispersion is measured with respect to a best-fit EW$_{\rm Li}$-$T_{\rm eff}$ isochrone that is already tuned for each cluster; the range of metallicities for the young GES clusters we have used, listed in table~3 of \cite{Randich2022a}, is only $-0.1<$ [Fe/H] $<+0.03$; and the trend of $\langle$ EW$_{\rm Li} \rangle$ vs age (see Fig.~\ref{fig3}) shows little evidence for any scatter that might explain the growing dispersion seen in Fig.~\ref{fig3}.

\end{itemize}

\subsection{Modelled lithium equivalent widths}
\label{model_ew}
EW$_{\rm m}$ was estimated as a function of age from model Li abundances over a grid of 17 evenly spaced  masses between 0.2 and 1\,M$_{\odot}$ and for five $\beta$ values between 0 and 0.4. The available curves of growth use 1D LTE atmospheres. Assuming the evolutionary models predict the 3D NLTE Li abundance, this is reverse-corrected to a corresponding 1D LTE abundance using the  {\sc breidablik} code of \cite{Wang2021a}, then folded through the LTE curves of growth (COG) presented in \cite{Franciosini2022b} to determine EW$_{\rm Li}$ at the temperature and $\log g$ of the spotted and unspotted surfaces (calculated assuming $x_{\rm spot}=0.8$). A weighted average was taken according to the relative flux contribution of the spotted and unspotted surfaces to give EW$_{\rm m}$ as a function of age, mass and spot coverage.

Neither the NLTE corrections or COGs extend to temperatures below 4000\,K, or at least not in a way that can be used to compare with our measured EW$_{\rm Li}$\footnote{The Franciosini et al. curves of growth predict pseudo-equivalent widths that contain contributions from blends and molecular features in stars with $T_{\rm eff}<4000$\,K, that are non-zero even in the absence of Li.}, so we are forced to make some approximations and accept that EW$_{\rm m}$ for cooler stars will have systematic error. For the NLTE  corrections we used the value for 4000\,K at lower temperatures. For the COGs we extrapolated curves of  EW/$\log{({\rm Li/Li}_0})$ versus $T_{\rm eff}$ to temperatures $<4000$\,K. The systematic errors will not be large for ZAMS K-dwarfs since their unspotted surfaces have $T_{\rm eff}>4000$\,K and the contribution from the spotted surface is suppressed by a factor $\beta x_{\rm spot}^4/(1 - \beta + x_{\rm spot}^4) \leq 0.16$ (for $\beta \leq 0.4$) with respect to the unspotted surface. More uncertainty is expected for the M-dwarfs where the unspotted surface has $T_{\rm eff} < 4000$\,K.

The curves of growth are summarised in Fig.~\ref{figA4}. These show that the NLTE corrections to EW$_{\rm m}$ are positive over most of the relevant $A$(Li)-$T_{\rm eff}$ domain. They also illustrate that the highest EW$_{\rm Li}$ values measured in a young (and presumably undepleted) cluster are consistent with an initial NLTE abundance $A$(Li)$\simeq 3.3$.

\begin{figure}
   \includegraphics[width = 85mm]{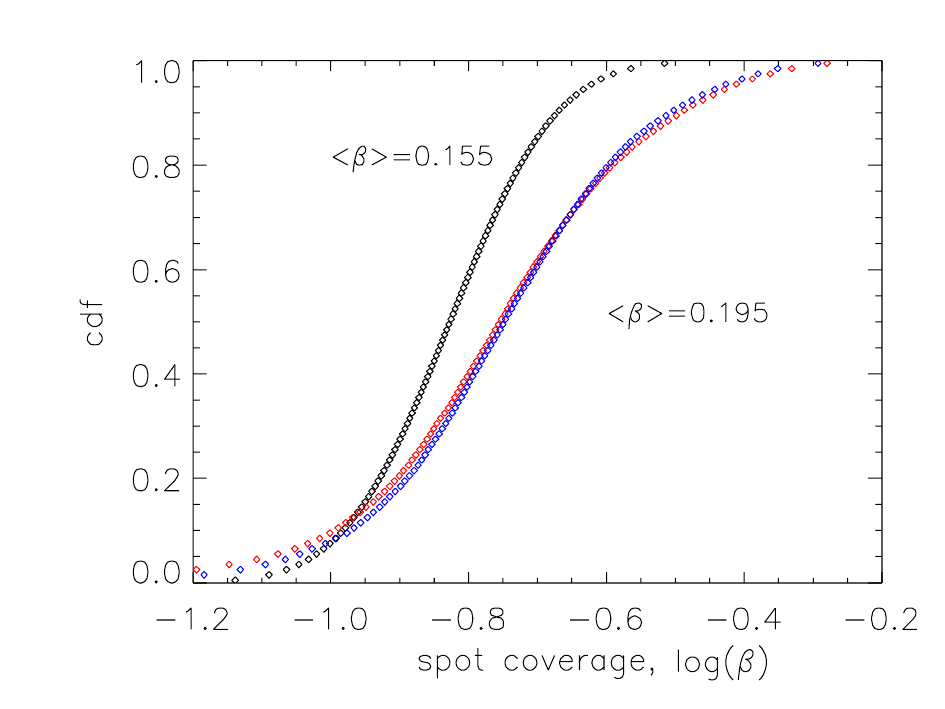}
   \caption{Cumulative probability distribution of flux blocking factor, $\beta$, used to calculate the model EW$_{\rm Li}$ dispersion. The left hand curve (in black) shows the CDF for a normal distribution in $\log \beta$ with a mean of -0.822$\pm$0.121\,dex (see \S\ref{Pleiades beta}). The curves to the right show the CDF for a more complex distribution comprising a normal distribution of $\log \beta$ convolved with a distribution of the average level of $\log \beta$ that reproduces the measured variation in $\delta$EW$_{\rm Li}$ with $\log$\,Period (see \S\ref{spot_rotation}). The very similar blue and red curves use the {\sc spots} and Pisa model calculations respectively.}
	\label{fig7}	
\end{figure}

\begin{figure*}
	\begin{minipage}[t]{0.98\textwidth}
            \includegraphics[width = 170mm]{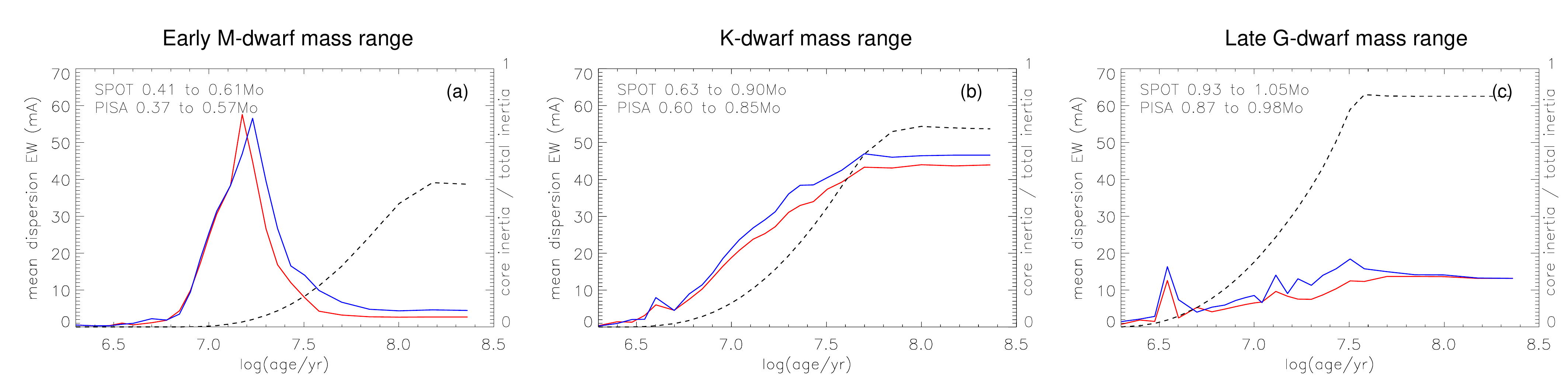}
        \end{minipage}
	\caption{Calculated levels of dispersion in Lithium equivalent width, $\Delta$EW$_{\rm m}$ as a function of age using tabulated data from the {\sc spots}  \citep{Cao2022a}~and Pisa \citep{Tognelli2021a} evolutionary models. Results are averaged over 3 mass ranges representing M, K and G-dwarfs (see \S \ref{mass_ranges}). Blue curves use {\sc spots} data and red curves Pisa data both assuming a Normal distribution of spot coverage $\log{\beta}=-0.822\pm 0.121$. The dashed line and the right hand axis shows the ratio of the {\sc spots} model core inertia to the total stellar inertia as a function of age.}
	\label{fig8}	
\end{figure*}

\subsection{A Pleiades-based model of spot coverage}
\label{spot_coverage}
\label{Pleiades beta}

\begin{table}
\caption{Estimated proportion of ZAMS K-dwarfs in the saturated regime ($\log$N$_{\rm R}< -0.677$) as a function of age.}
\begin{tabular}{lccc}
\hline
Fiducial &  \multicolumn{2}{c}{Numbers of targets} & Percentage with\\
age range   & Saturated   &  Unsaturated & $\log$N$_{\rm R}< -0.677$ \\\hline
$<$13\,Myr         &  192 &   1 &  99\% \\
13 to 25\,Myr      &   65 &   1 &  98\% \\
25 to 75\,Myr      &   72 &  17 &  80\% \\
75 to 250\,Myr     &  115 & 173 &  39\% \\
\hline
\end{tabular}
\label{saturation}
\end{table}

Modelling the dispersion in EW$_{\rm m}$ requires an estimate of the spot coverage and its dispersion as a function of temperature and age for PMS stars. There is little direct data available on this but as a first attempt we adopted the distribution of spot properties ($f_{\rm spot}$, $x_{\rm spot}$) inferred for low-mass stars in the ZAMS-age Pleiades cluster from modelling their spectra with two-temperature atmospheres \citep{Fang2016a, Cao2022a}.

\cite{Cao2022a} found that $f_{\rm spot}$ saturates for the most rapid rotators with small Rossby numbers ($N_{\rm R}$, the ratio of period to convective turnover time), with a decline at slower rotation rates and larger Rossby numbers. Using results from table~1 of \cite{Cao2022a} we find $\beta$ has a roughly lognormal distribution with a mean $\log{\beta } = -1.066 \pm 0.365$ if considering all single K/M-dwarfs in the Pleiades, and a mean $\log{\beta} = -0.822 \pm 0.121$ for stars with $\log{N_{\rm R}}< -0.677$ in the saturated regime. The quoted error bars are standard deviations and we assumed these represent a genuine, long-lived dispersion in spot coverage. 

In this model, the scatter in $\beta$ between targets at a given age would thus strongly depend on the proportion of stars that were in the saturated regime. To establish this for our sample clusters, the $v \sin i$ in Table 1 was used to estimate $N_{\rm R}$ as the ratio of the expectation value of rotational period $P_{\rm rot}$ from Eqn.~\ref{Prot}, to a turnover time taken from the appropriate {\sc spots} model. Table \ref{saturation}~shows the fraction of stars in the ZAMS K-dwarf sample that are in the saturated regime in the different age bands. At ages $<75$\,Myr, where the {\sc spots} (and Pisa) models predict the vast majority of PMS Li-burning takes place, $\sim 90$ per cent of ZAMS K-dwarf targets (and all of the ZAMS early M-dwarfs) are in the saturated regime. On this basis, for the purpose of predicting Li depletion during and after PMS Li burning, we assume a simple, age-independent, normal distribution in $\log \beta$ with an average of $-0.822$ and 1-sigma dispersion of 0.121, capped at a maximum $\beta \le 0.47$ (equivalent to $f_{\rm spot} \le 0.8$ for $x_{\rm spot} = 0.8$). 

Using the {\sc spots} models of Li depletion as a function of age, mass and $\beta$ (Fig.~\ref{fig5}), and the NLTE corrections and COG  previously described, EW$_{\rm m}$ was calculated for 100 points over the cumulative distribution function (CDF) of $\log{\beta}$ (the black curve in Fig.~\ref{fig7}) and the results averaged to give the the mean and dispersion of EW$_{\rm m}$ as a function of age for 50 masses over the full mass range. These results were then averaged to give the dispersion $\Delta$EW$_{\rm m}$ as a function of age over the relevant ZAMS G-, K- or M-dwarf mass range.  The process was repeated using the Pisa models of Li depletion.

The results, shown in Fig.~\ref{fig8}, have a very similar time-dependence to the observed $\Delta {\rm EW}_{\rm Li}$ versus age relationships in Figs.~\ref{fig3} and~\ref{fig4}, although the maximum values of $\Delta $EW$_{\rm m}$ are only about half of the observed values. The results using the {\sc spots} and Pisa evolutionary models are also very similar. Figure~\ref{fig8}a shows a transient peak in $\Delta $EW$_{\rm m}$ for the early M-dwarf mass range that falls to zero before stars reach the main sequence. Figure~\ref{fig8}b shows a steady increase in $\Delta$EW$_{\rm m}$ over the epoch of PMS Li-burning for the K-dwarf range that persists up to the main sequence and Fig.~\ref{fig8}c, representing ZAMS G-dwarfs,  shows only a small $\Delta$EW$_{\rm m}$ over the full age range. One other significant difference between the model and observed dispersion is that there is no predicted change in $\Delta$EW$_{\rm m}$ for K-type stars $>75$ Myr. This is because the {\sc spots} (and Pisa) models do not feature any non-convective mixing, and predict little change in Li abundance after the initial PMS phase of burning (see Fig.~\ref{fig6}). However, the cluster observations show that the Li abundance of the K-stars does decrease $>75$ Myr (see Fig~\ref{fig3}c).

\subsection{Sensitivity to assumed model parameters}

Figure~\ref{fig8} was calculated for "warm spots" with $x_{\rm spot}=0.8$. Calculations were repeated for two other cases, "dark spots" ($x_{\rm spot}=0$) where the spotted surface makes no contribution to stellar luminosity and a "grey surface" where the surface temperature of a star of a given luminosity is reduced by a uniform factor of $(1-\beta)^{1/4}$, representing the case where stellar radii are inflated (at a fixed mass and age) by uniform changes in stellar structure \citep{Feiden2014a} rather than darker spots. For the early M- and K-dwarf ranges, the $\Delta$EW$_{\rm m}$ results for dark spots and a grey surface differ by a negligible $<$5\,m\AA~from the warm spots case, with up to twice this difference being seen in the late G-dwarf mass range.

\subsection{A rotation-dependent spot model}

\label{spot_rotation}

The model for spot coverage and spot filling factor dispersion in \S\ref{spot_coverage}, based on observations of K and M-stars in the Pleiades, fails by a factor of two to explain the magnitude of the observed Li dispersion. As formulated, it also does not inject any rotation-dependence into the modelled Li depletion and could not explain the strong correlations seen in Fig.~\ref{fig6}. To address these shortcomings a more speculative spot model was considered that allows $\beta$ to increase with rotation rate, even if the stars have rotation rates that put them well below the conventional threshold Rossby number for magnetic saturation. 
Since the evolutionary models do not allow a time-dependent spot-coverage, we are forced to assume that the rotation-dependent $\beta$ for any star applies throughout the epoch of Li-burning after which, the Li dispersion is fixed. 

 Taking all the targets in the ZAMS K-dwarf mass range from Fig.~\ref{fig6} in the age range 12 to 75\,Myr, the epoch of PMS lithium burning, a linear least squares fit gives 
 \begin{equation}
 \langle \delta{\rm EW}_{\rm Li} \rangle = 26-95\,\log({\rm period/d})\ ,
 \end{equation}
 where $\delta{\rm EW}_{\rm Li}$ is the deviation from the mean relationship defined by an {\sc eagles} isochrone in m\AA. If this linear trend were the sole cause of star-to-star EW$_{\rm Li}$ differences it would yield $\Delta$EW$_{\rm m} \simeq 43$\,m\AA, indicating that there must still be additional scatter perhaps due to a spread in $\beta$ at a given rotation rate.

\begin{figure}
            \includegraphics[width = 70mm]{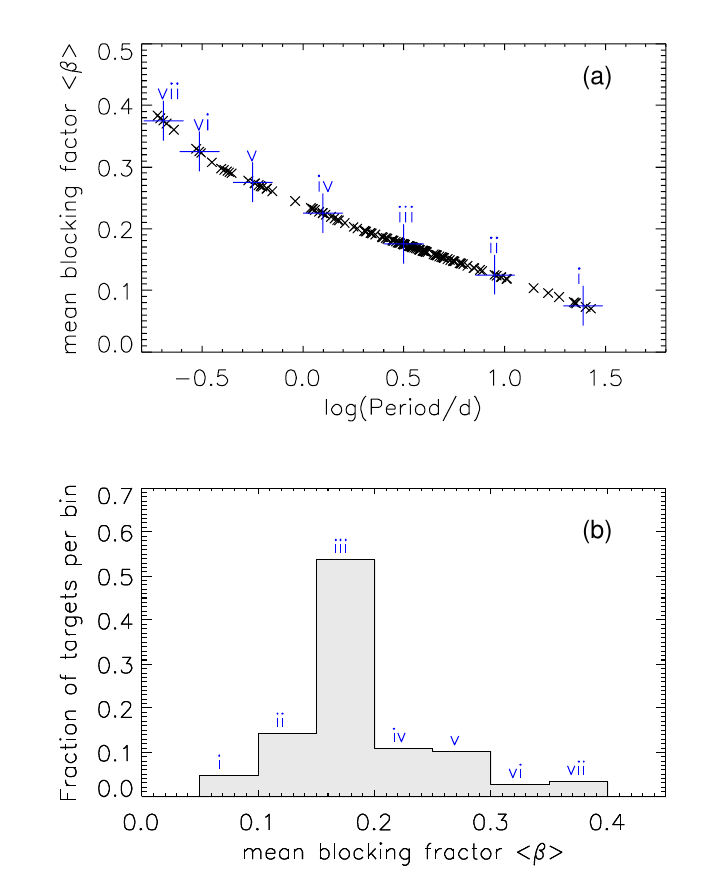}
	\caption{(a) The relationship between mean flux blocking factor $\langle \beta \rangle$ and rotation period required to produce the corresponding observed correlation between the offset in EW$_{\rm Li}$ from an {\sc eagles}-predicted isochrone versus rotation period, for targets in the ZAMS K-dwarf mass range with ages 12--75\,Myr. (b) The normalised distribution of $\langle \beta \rangle$ for GES cluster targets in this mass and age range after applying the relationship in (a) to their estimated rotation periods. Roman numerals on plot a mark the rotational periods corresponding to the mean $\langle \beta \rangle$ of the bins in plot b.}
\label{fig10}	
\end{figure}

\begin{figure*}
	\begin{minipage}[h]{0.98\textwidth}
 \centering
            \includegraphics[width = 140mm]{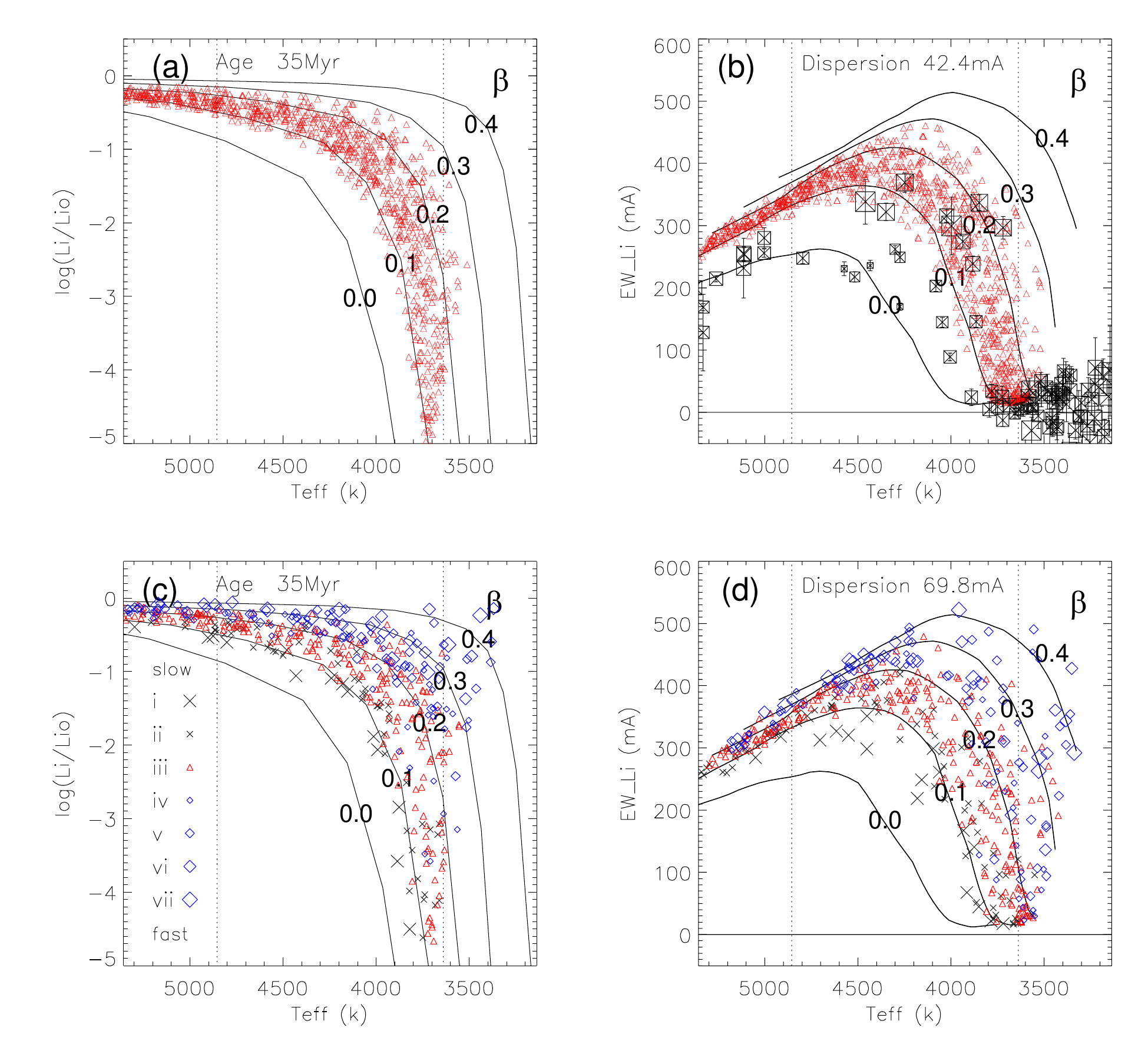}
     \end{minipage}
	\caption{Plots a and b show predicted values of $\log{{\rm Li/Li_0}}$ and EW$_{\rm m}$ as a function of $T_{\rm eff}$ using the {\sc spots} model. Black lines show results at increasing levels of $\beta$. Red triangles show random results drawn from a Normal distribution of $\log{\beta}=-0.822\pm 0.121$. Vertical dashed lines show the limits of the K-dwarf mass range at the model age of 35\,Myr. Black points on plot b show measured levels of EW$_{\rm Li}$ in NGC\,2547. Plots c and d show model levels of EW$_{\rm m}$ drawn from $\log$~Normal distributions at 7 levels of $\langle \beta \rangle$, increasing with rotation rate, with a dispersion in $\log \beta$ of $\sigma$=0.121, with the number of points plotted weighted according to the likelihood of $\langle \beta \rangle$. Points are shown with symbols corresponding to the rotation-dependent categories indicated with Roman numerals (i-vii), corresponding to the same numerals in Fig.\ref{fig10}. These categorise points by decreasing rotation period, and hence increasing $\langle \beta \rangle$.}   
	\label{figA5}	
\end{figure*}

The evolutionary models can be used to estimate by how much $\langle \beta \rangle$ would have to change with rotation, with respect to an assumed overall sample mean value, in order to reproduce the observed dependence of EW$_{\rm Li}$ on rotation, when averaged over the specified range in mass and age. Figure~\ref{fig10}a shows an example of this calculation for an assumed overall mean $\langle \beta \rangle = 0.2$ and Fig.~\ref{fig10}b shows how this relationship with rotation translates into a distribution of $\langle \beta \rangle$ for the observed targets in the 12-75\,Myr sample. This distribution was then convolved with the standard deviation of $\log \beta$ observed in the saturated stars of the Pleiades (0.121 dex) and Fig.~\ref{fig7} shows the resulting cumulative distribution function of $\log \beta$. A similar calculation using the Pisa models yields an almost identical distribution. 

As an example of how the process works Fig.~\ref{figA5}  shows a simulated population of stars at 35\,Myr in both the abundance and EW$_{\rm Li}$ versus $T_{\rm eff}$ planes for the spot model with no rotation dependence from \S\ref{spot_coverage} and for the rotation-dependent spot model. Lines marking isochrones for a given $\beta$ are shown and the results for the rotation-dependent simulation show where simulated stars with a range of rotation periods (according to the labels in Fig.~\ref{fig10}) end up. A comparison with the data for NGC~2547 (age $\simeq 35$ Myr) shows that the dispersion is under-predicted by the spot distribution with no rotation dependence, reasonably reproduced by the rotation-dependent model thanks to the greater spread in $\beta$, but that both models over-predict the observed mean level of lithium in NGC~2547.

\begin{figure}
    \centering
    \includegraphics[width=85mm]{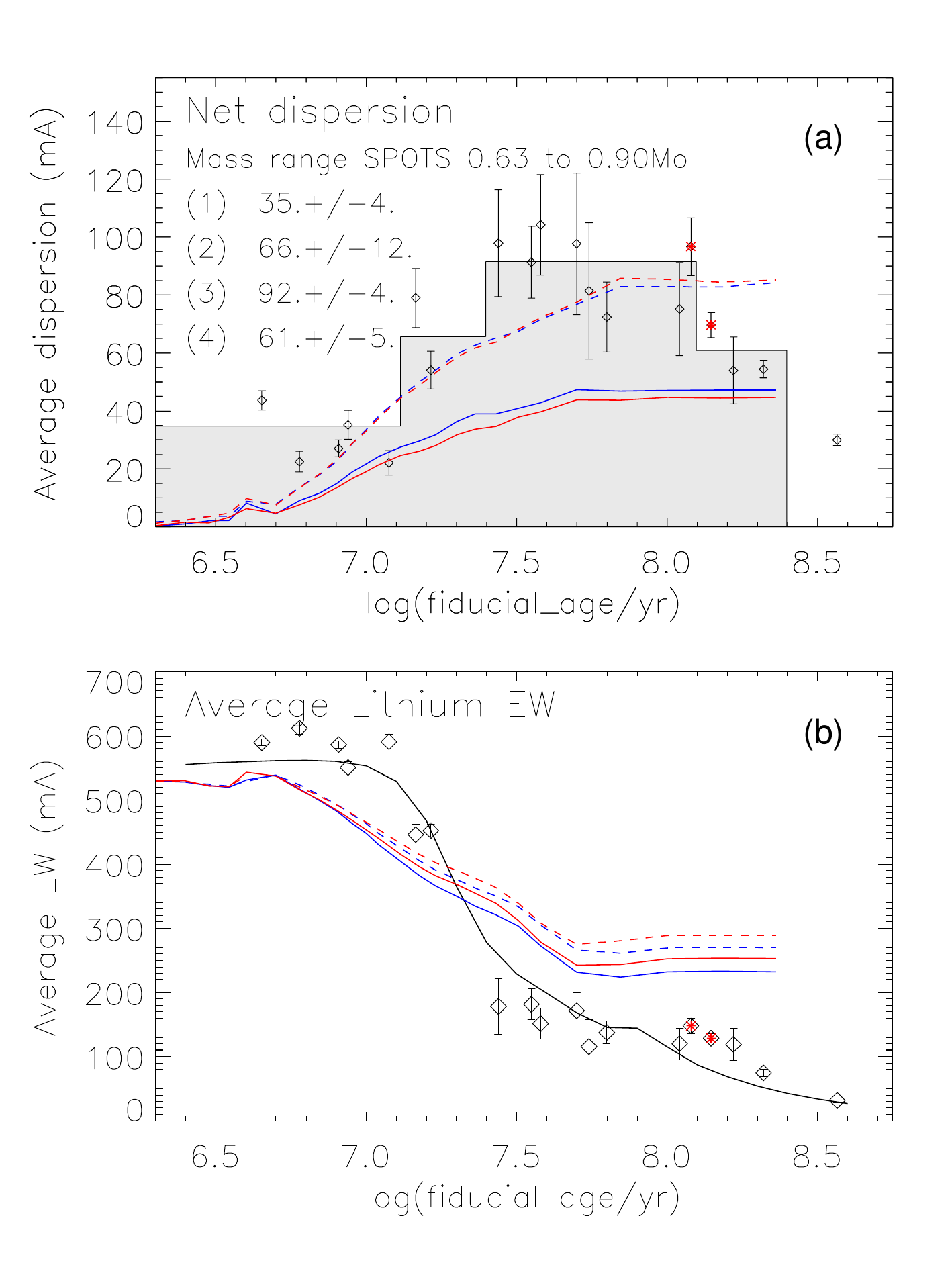}
    \caption{Plot a shows the measured dispersion in EW$_{\rm Li}$ for the ZAMS K-dwarf mass range (as Fig.~\ref{fig3}b) compared with the modelled dispersion for the starspot distributions based on the Pleiades measurements (solid lines, \S\ref{Pleiades beta}) and the rotation-dependent starspot distribution (dashed lines, \S\ref{spot_rotation}). Plot b shows average levels of  EW$_{\rm Li}$ (as Fig.~\ref{fig3}c) compared to model average values.  Red and blue curves show results calculated using the Pisa and {\sc spots} evolutionary models respectively.}
    \label{fig11}
\end{figure}

Assuming this $\beta$ distribution applies at all ages during Li-burning, the methods described in \S\ref{spot_coverage} were used to calculate $\Delta$EW$_{\rm m}$ as a function of age. $\Delta$EW$_{\rm m}$ versus age for ZAMS K-dwarfs using this rotation-dependent model is compared with the GES data and with the simple saturated model from \S\ref{Pleiades beta} in Fig.\ref{fig11}. The temporal behaviour of the predicted Li dispersion is very similar to the saturated model but, by design, the extra rotation-dependence injects significantly more dispersion at all ages and nearly enough to match the data.

The agreement between $\langle {\rm EW}_{\rm m} \rangle$ and  $\langle {\rm EW}_{\rm Li} \rangle$ is less good. This appears to be because the model already predicts significant Li depletion after 10\,Myr, which is not reflected in the observations. This could be an indication that the average $\beta$ is even higher than we have assumed at the youngest ages, since higher $\beta$ models begin depletion later (see Fig.~\ref{fig5}). Secondly, the rate of depletion after 10\,Myr seems too slow in the models, and Li depletion stops altogether in the models after around 75\,Myr, when these stars reach the ZAMS. The observations show that it continues, suggesting that additional non-convective mixing mechanisms, that are not included in these evolutionary models, must be at work.

Further tinkering with the dispersion in $\log \beta$ or with the assumed overall mean $\langle \beta \rangle$ might achieve a better agreement, as might the use of evolutionary models that allow $\beta$ to change with age and which include additional non-convective mixing mechanisms. However, from this test case we can conclude that spot coverage would need to systematically vary by a factor of several between rotation periods of 0.3 and 10 days (see Fig.~\ref{fig10}a) during the Li-burning phase in order to reproduce the observed magnitude of $\Delta$EW$_{\rm Li}$ and the EW$_{\rm Li}$-rotation correlation. 
If spot coverage does not vary with rotation for low-mass PMS stars at ages of 12-75\,Myr then variable spot coverage cannot be the driver of the observed Li-rotation connection.  

\section{Discussion}
\label{sec:discussion}

The results in \S3 show that the dispersion in EW$_{\rm Li}$ ($\Delta$EW$_{\rm Li}$) first develops in low-mass stars between 10 and 20\,Myr, coincident with the onset of Li depletion. For stars in the mass range that will become K-dwarfs at 120\,Myr, this also coincides with the growth of a radiative core, according to the {\sc spots} models. For stars that become M dwarfs at 120 Myr, Li depletion and dispersion is already well established {\it before} the radiative core forms (Figs.~\ref{fig4} and~\ref{fig9}). At least for these latter, lower-mass stars, this argues against mechanisms to explain the dispersion that involve rotation-dependent overshooting at the convection zone base \citep{Baraffe2017a} or additional mixing associated with rotational shear and core-envelope decoupling \citep{Bouvier2008a}. Of course, these observations do not rule out that such mechanisms might play some role in higher mass stars once the radiative core has developed.

A caveat that should be mentioned here is that to estimate the EW$_{\rm Li}$ dispersions we are forced to combine data for stars in bins of $T_{\rm eff}$ and age, and in some cases, for example, the late K- and M-dwarfs, Li depletion is predicted to be rapid and a strong function of $T_{\rm eff}$. Our methods described in \S\ref{measured_dispersion} have attempted to account for this by including dispersion due to $T_{\rm eff}$ uncertainties and by plotting the results for individual clusters at their estimated ages. Nevertheless the study could certainly be improved by the inclusion of even more data from different clusters in order to provide better resolution in $T_{\rm eff}$ and age.

In \S\ref{spot_model} we investigated whether the dispersion could be explained in terms of a variable coverage of cool starspots during the epoch of lithium destruction, using the observed variation of spot coverage among Pleiades K/M dwarfs as a template. The results, summarised in Fig.~\ref{fig8} and compared with Figs.~\ref{fig3} and~\ref{fig4}, show that the temporal evolution of $\Delta$EW$_{\rm Li}$ is captured well by a simple two-temperature atmosphere model with a variable spot coverage, irrespective of the timing of the development of the radiative core, for stars that will become G, K or M stars at 120 Myr. A caveat here is that the conversion from model abundances to observed EW$_{\rm Li}$ is likely more complex than simulated by a simple two-temperature atmosphere folded through a curve of growth.

There are two weaknesses in a simple spot hypothesis for the Li-dispersion. The first is that modelling spot coverage, and dispersion using the observed results for Pleiades stars at 120\,Myr \citep{Cao2022a}, yields a $\Delta$EW$_{\rm m}$ that is a factor of two smaller than that observed for clusters at 25-125\,yr. This is despite the generous assumption that the spread in $\beta$ inferred by Cao et al. is due to variations in long-term spot coverage. In fact, this must be an upper limit after considering observational uncertainties and short-term changes in spot coverage. Such changes, associated with asymmetric spot distributions or activity cycles on decadal timescales, that may change $\beta$ by $\sim \pm 0.1$, have been seen in ZAMS K-dwarfs \citep[e.g.,][]{Innis1988a, Jarvinen2005a}. These would not cause significant variations in the long-term evolution and Li-depletion of the star but would cause additional scatter in instantaneous measurements of starspot coverage.

Secondly, since we judged that most of our sample stars would have $\log N_R \lesssim -0.7$ and saturated levels of magnetic activity during the epoch of Li-burning, this lead us to assume a constant mean and relatively narrow spread in $\beta$. This in turn limits the predicted $\Delta$EW$_{\rm Li}$ and predicts no correlation at all between EW$_{\rm Li}$ and rotation as the dispersion develops. In contrast, in \S\ref{li_rotation}, we showed that the dispersion in EW$_{\rm Li}$ is strongly correlated with rotation in  stars that will become ZAMS K-dwarfs almost as soon as Li depletion commences at 10-20\,Myr. This means either that spot coverage has a higher dispersion and rotation-dependence in PMS stars much younger than the Pleiades or that there is still some other rotation-dependent  process affecting the Li depletion in the K-dwarf mass range that causes the correlations so clearly seen in Fig.~\ref{fig6}.

If spots are to drive the Li-rotation correlation, then in \S\ref{spot_rotation} we showed that if $\langle \beta \rangle \simeq 0.2$ in PMS stars during the epoch of Li burning (12-75 Myr), there must also be a rotation dependence of $\beta$, such that $\langle \beta \rangle$ for stars with the shortest estimated rotation periods ($\sim 0.3$ days) must be a few times larger than for the slowest rotators with periods $\sim 10$ days. This rotation dependence is needed to explain the observed correlation of EW$_{\rm Li}$ with rotation period seen at the ZAMS. However, since all but the very slowest rotators in this mass range would have $\log N_R < -0.7$ (see Table~\ref{saturation}) the vast majority of the population should have saturated levels of magnetic activity. By that, we mean numerous observations of proxy magnetic activity indicators such as chromospheric emission and coronal X-ray fluxes have been observed to increase with decreasing Rossby number up to about this threshold and then flatten off for faster rotating stars (\citealp*{Noyes1984a}, \citealp{Vilhu1984a}, \citealp{Pizzolato2003a}, \citealp{Wright2011a}). This saturation of magnetic activity indicators has also been observed in samples that include PMS stars at 12-75 Myr in the mass ranges we consider (e.g., \citealp*{Marsden2009a}, \citealp{Jeffries2011a}).

Whether starspot coverage or the general level of magnetic flux at the photosphere obey a similar relationship with Rossby number is more poorly defined. Although the data are scattered, neither \cite{Fang2016a} or \cite{Cao2022a} present evidence that could be consistent with spot coverage in ZAMS K- and M-dwarfs growing by the necessary amounts at low Rossby numbers. However, Zeeman broadening and intensification of spectral lines has been used to show that the mean magnetic field does continue to increase as $N_R^{-0.2}$ in ``saturated" Pleiades M-dwarfs \citep{Wanderley2024a}, and a similar rate of increase is also observed with decreasing period \citep[see their fig.~4, which also includes data for fast rotating M-dwarfs from][]{Shulyak2017a, Shulyak2019a, Reiners2022a}. This leads to an approximate doubling of the mean magnetic field and quadrupling of the magnetic pressure contribution between rotation periods of 10 days to the fastest rotators in the Pleiades.

There are also many reports for PMS stars with much higher spot coverage than inferred for the Pleiades ZAMS K-dwarfs. A number of studies have shown that there is much better agreement with colour-magnitude diagrams and the overall Li depletion patterns of PMS clusters for models featuring levels of spot coverage with $\beta \geq 0.2$ \citep{Jeffries2017a, Binks2021a, Binks2022a, Franciosini2022a}. Photometric monitoring and modelling of variable young weak-lined T-Tauri stars (WTTS) in the Taurus-Auriga association (ages of a few Myr) suggest $0.2 < \beta < 0.5$ in those sources with the largest variability amplitudes \citep{Grankin1999a, Grankin2008a}.
Spectroscopic investigations, similar to those performed in the Pleiades, fitting two temperature components to Ti\,O bands in the optical spectrum yielded $\beta \simeq 0.4$ in the K-type WTTS V410 Tau \citep[rotation period 1.9 days,][]{Petrov1994a}. Two-temperature analyses of near-IR spectra in Taurus-Auriga suggest $0.33 < \beta < 0.68$ in a sample of 10 K- and M-type WTTS, with rotation periods of 2--7 days, but with no significant correlation between $\beta$ and rotation period \citep{GullySantiago2017a, PerezPaolino2024a}.

The overall level and rotation dependence of starspot filling factors in PMS stars younger than the Pleiades still seems to be an open question but is key to ruling in or out a starspot explanation for the observed level and rotation dependence of the Li dispersion. It is of course also possible that while simple starspots could provide an explanation for some of the observed dispersion, there may still be significant contributions to a dispersion in EW$_{\rm Li}$ from plages, chromospheric activity, overionisation and other NLTE effects that may themselves be rotation-dependent \citep{Stuik1997a}. Disentangling these effects from abundance variations might be done using the K\,{\sc i}~7699\,\AA\ analogue of the Li\,{\sc i}~6708\,\AA\ line relied on here, since star-to-star K abundance variations are unlikely.

\section{Conclusions}

Lithium is not uniformly depleted during the PMS phase in low-mass stars. A large dispersion in Li abundances and the corresponding Li~{\sc i}~6708\AA\ equivalent widths is known to exist in the K-dwarfs of ZAMS clusters like the Pleiades and is strongly correlated with rotation rate; a result that is not predicted by standard evolutionary models. Using a large sample of homogeneously determined Li~{\sc i}~6708\AA\ equivalent widths, the development of this dispersion in Li at a given age and $T_{\rm eff}$ is explored for PMS and ZAMS stars in 19 open clusters between 2--300\,Myr, mostly from spectra obtained as part of the {\it Gaia}-ESO spectroscopic survey. The main observational conclusions are that:
\begin{itemize}
    \item The dispersion begins at 10--20\,Myr in stars destined to be K- or M-dwarfs at the ZAMS, coinciding with the initiation of Li depletion.

    \item The dispersion begins alongside the growth of a radiative core in what will be ZAMS K-dwarfs, but becomes well-established in lower mass stars even whilst they are fully convective.

    \item The correlation between higher EW$_{\rm Li}$ and faster rotation rates also becomes strong and statistically significant between 13 and 25\,Myr.
\end{itemize}

Since the dispersion commences even in fully convective PMS stars it suggests, at least for stars of low mass, that mechanisms involving rotation-dependent, non-convective mixing, shear at the interface between a radiative core and convective envelope, or rotation-dependent convective overshoot into the core, are not responsible for this dispersion.

A model is investigated based on a variable coverage by dark, magnetic starspots, which are known to be present in PMS/ZAMS stars. This could lead to varying rates of Li depletion and therefore a dispersion in Li abundance and Li equivalent width at a given age and $T_{\rm eff}$. 

Adopting a distribution of starspot coverage inferred for ZAMS K- and M-dwarfs in the Pleiades leads to a predicted Li dispersion that matches the observations in terms of its temporal behaviour, but fails by a factor of two to provide enough dispersion and also predicts no correlation with rotation. Instead, a rotation-dependent spot coverage is introduced and it is shown that if the spot coverage during the phase of PMS Li depletion is larger on average than in the Pleiades {\it and} increases by factors of a few between the slowest and fastest rotators in our sample, then both the temporal evolution and the magnitude of the Li dispersion might be reproduced.

A difficulty for this model is that although there is some evidence that starspot coverage is greater in younger PMS stars, there is no direct evidence that it increases with rotation rate and such a variation would contradict the saturation that is observed at fast rotation rates in some other magnetic activity indicators.
There is an urgent need to establish the extent and rotation dependence of starspot coverage in PMS stars during the epoch of Li depletion, preferably in a comparable way to that achieved for ZAMS stars in the Pleiades.

\section*{Acknowledgements}
This work is based on data products from observations made with ESO Telescopes at the La Silla Paranal Observatory under programme IDs 188.B-3002, 193.B-0936 and 197.B-1074.
These data products were processed by the Cambridge Astronomy Survey Unit (CASU) at the Institute of Astronomy, University of Cambridge, and by the FLAMES/UVES reduction team at INAF/Osservatorio Astrofisico di Arcetri. 

This work has made use of data from the European Space Agency (ESA) mission
{\it Gaia} (\url{https://www.cosmos.esa.int/gaia}), processed by the {\it Gaia}
Data Processing and Analysis Consortium (DPAC,
\url{https://www.cosmos.esa.int/web/gaia/dpac/consortium}). Funding for the DPAC
has been provided by national institutions, in particular the institutions
participating in the {\it Gaia} Multilateral Agreement.

\section*{Data Availability Statement}

The parent catalogue \citep[from][]{Jeffries2023a} on which the analysis in this paper was based is available at \url{https://oup.silverchair-cdn.com/oup/backfile/Content_public/Journal/mnras/523/1/10.1093_mnras_stad1293/2/stad1293\_supplemental_files.zip}.
The reduced stacked spectra underlying that paper can be obtained via the European Southern Observatory (ESO) archive and is identified as the ‘Phase 3 release’ of Gaia-ESO spectroscopic survey DR4 : \url{https://archive.eso.org/scienceportal/home?data_collection=GAIAESO&publ_date=2020-12-09}.
The full catalogue of stellar parameters derived from these spectra by the various GES working groups is also available in the ‘Phase 3 release’ of the Gaia-ESO spectroscopic survey DR5.1 : \url{https://archive.eso.org/scienceportal/home?data_collection=GAIAESO&publ_date=2023-07-02} 
Raw data can also be obtained from the ESO archive.

The {\sc eagles} code is available at \url{https://github.com/robdjeff/eagles}.

For the purposes of open access, the author has applied a Creative Commons Attribution (CC-BY) licence to any Accepted Author Manuscript version arising from this submission.

\bibliographystyle{mnras}
\bibliography{Li_dispersion_FINAL} 

\appendix

\section{Supplementary plots}
\begin{figure*}
	\begin{minipage}[t]{0.98\textwidth}
        \centering
            \includegraphics[width = 150mm]{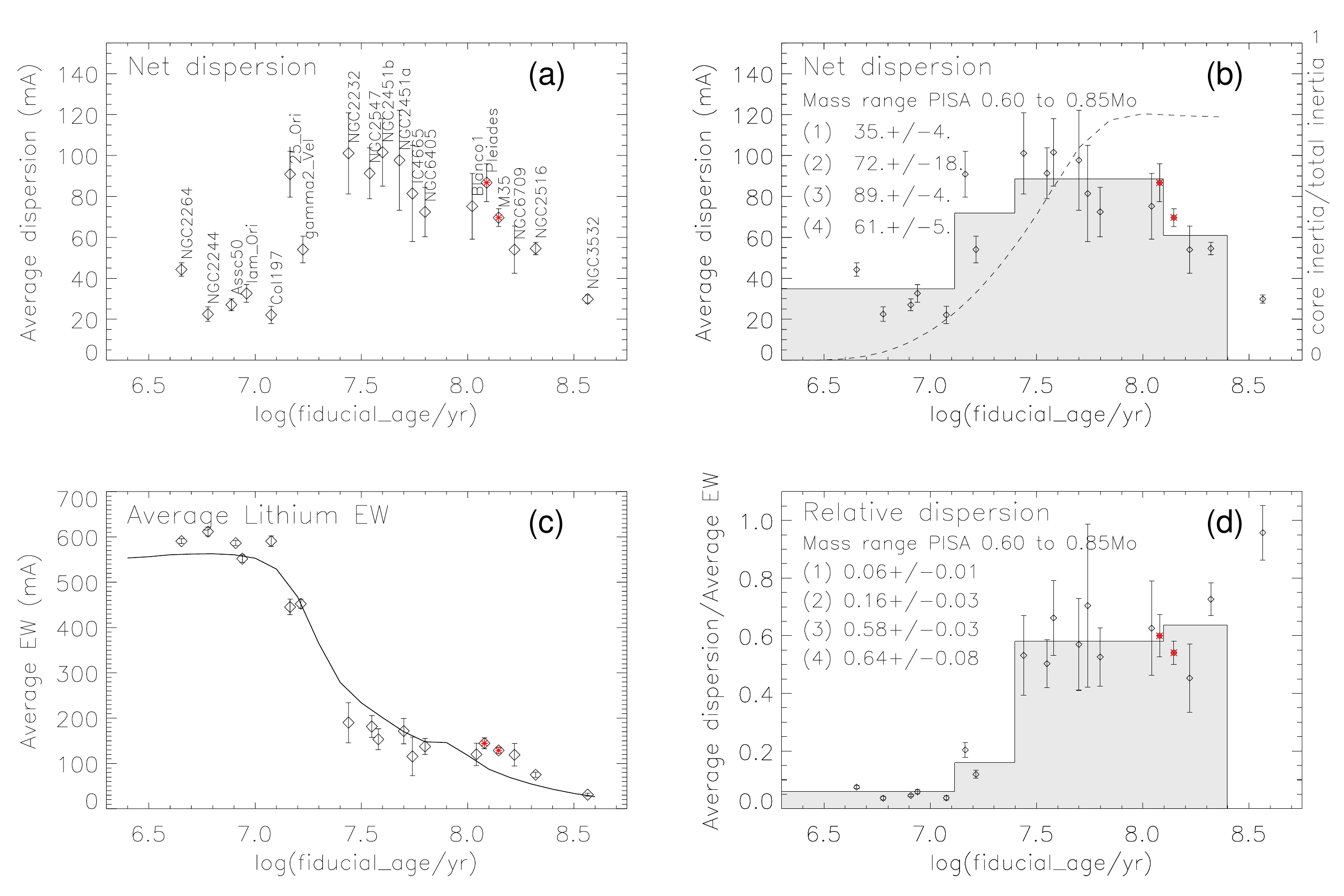}
        \end{minipage}
	\caption{The equivalent of Fig.~\ref{fig3} but using the Pisa stellar evolutionary models (instead of the {\sc spots} models) to define the $T_{\rm eff}$ range of stars that will become ZAMS K-dwarfs.}
	\label{figA1}	
\end{figure*}

\begin{figure*}
	\begin{minipage}[h]{0.98\textwidth}
 \centering
            \includegraphics[width = 150mm]{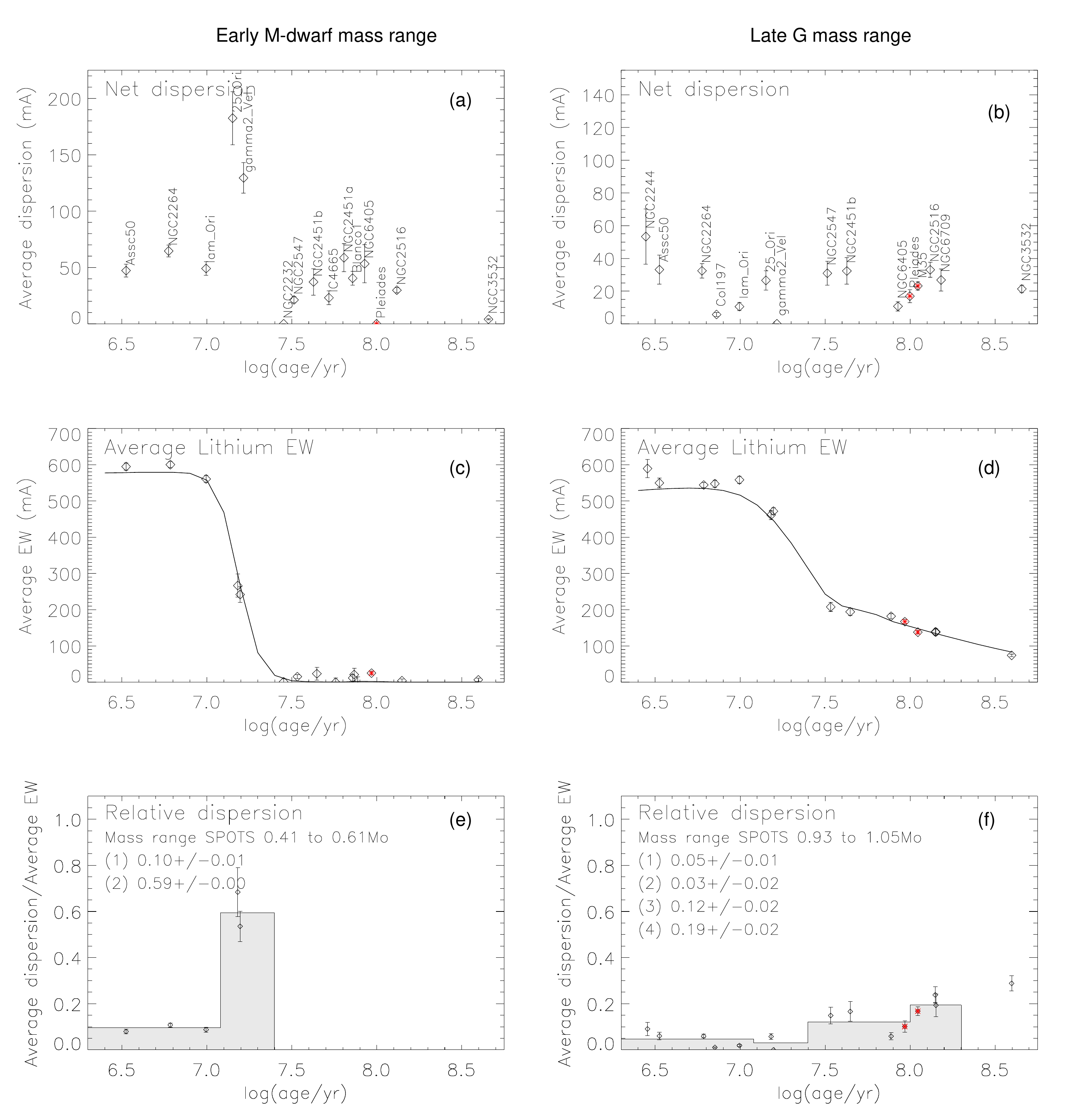}
     \end{minipage}
	\caption{EW$_{\rm Li}$ dispersion of stars in the early M (left hand plots) and late G-dwarf (right hand plots) mass ranges from the {\sc spots} models as a function of log age. Plots a and b  shows $\Delta$EW$_{\rm Li}$ (see Eqn.~\ref{Eqn1}) for individual clusters with small age offsets for clarity of cluster names. Plots c and d show average levels of EW$_{\rm Li}$ for individual clusters, $\langle{\rm EW}_{\rm Li}\rangle$, with the curve indicating the {\sc eagles} prediction averaged over the appropriate $T_{\rm eff}$ range. Plots e and f show the normalised relative dispersion, $\Delta$EW$_{\rm Li}/\langle{\rm EW}_{\rm Li}\rangle$ for individual clusters, with a histogram and text on the plot showing weighted average values in four age bands. The normalised relative dispersion is not defined for the early M-dwarf mass range at ages $>$25\,Myr since $\langle$EW$_{\rm Li}\rangle$ tends to zero for these stars and can be negative for individual clusters.}
	\label{figA2}	
\end{figure*}

\begin{figure}
   \includegraphics[width = 85mm]{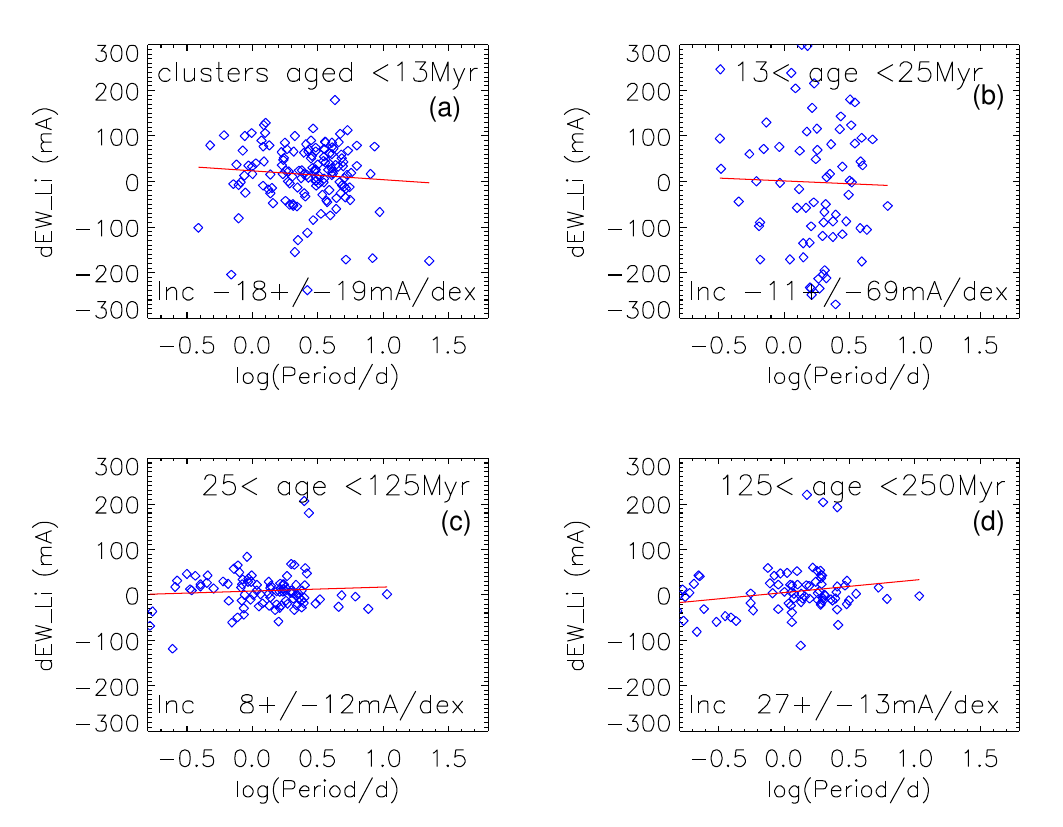}
   \caption{The offset of measured EW$_{\rm Li}$ for targets in the ZAMS early M-dwarf mass range (from the {\sc spots} models) with respect to the {\sc eagles} EW$_{\rm Li}$ isochrone, at the $T_{\rm eff}$ and age of the target, versus rotation period estimated from  $v\sin{i}$ (see \S\ref{li_rotation}). Red lines show a linear least square fit; text on the plots gives the best fit slope and uncertainty.}
	\label{figA3}	\
\end{figure}

\begin{figure}
   \includegraphics[width = 85mm]{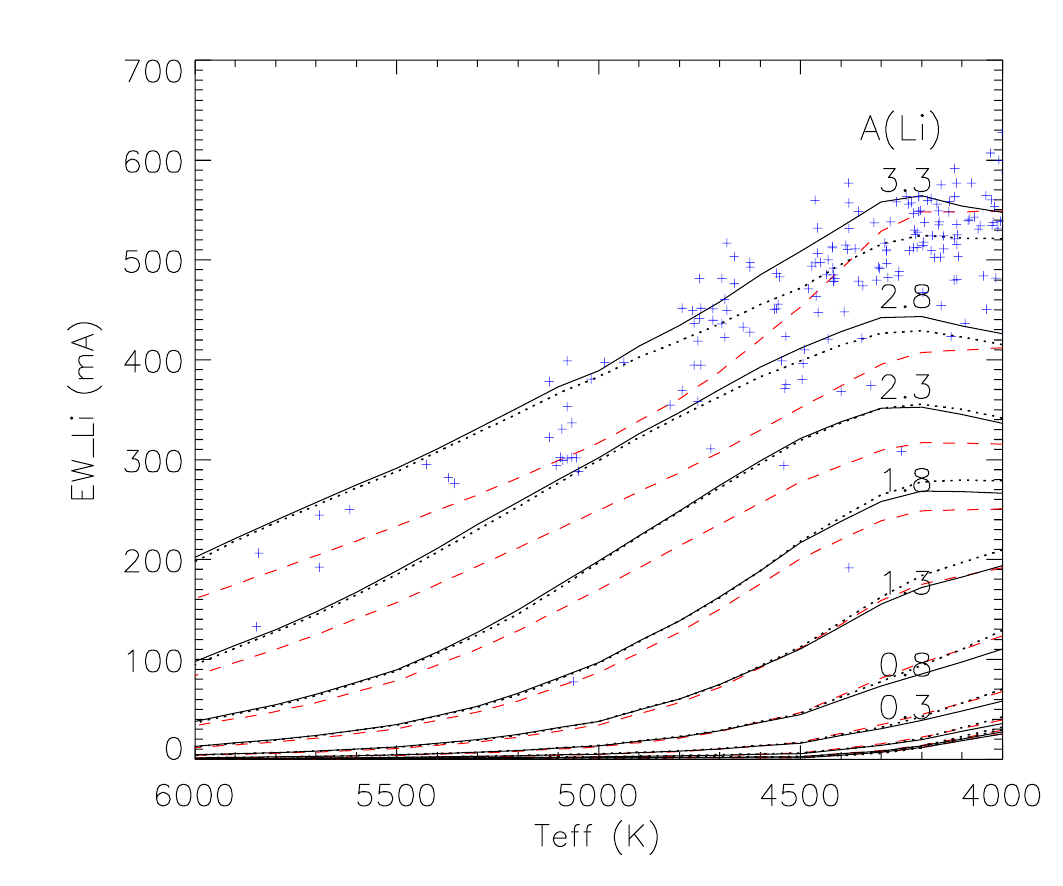}
   \caption{Curves of growth used to predict model values of EW$_{\rm Li}$ for a given $A$(Li) and $T_{\rm eff}$. Red dashed lines show the LTE curves from  \protect\cite{Franciosini2022b} for $\log g =4.5$ and black lines show the corresponding NLTE-corrected COGs using the code of \protect\cite{Wang2021a} (see \S\ref{model_ew}).  Dotted lines show the corresponding NLTE-corrected curves for $\log g=4.0$. Blue points show EW$_{\rm Li}$ measurements in the young cluster NGC\,2264.}
	\label{figA4}	
\end{figure}

\bsp 
\label{lastpage}
\end{document}